\documentclass[aps,prx,twocolumn,superscriptaddress]{revtex4-1}

\usepackage{bm}
\usepackage{braket}
\usepackage{amsmath}
\usepackage{amssymb}
\usepackage{mathtools}
\usepackage{dsfont}

\begin{document}

\preprint{}

\title{Observation of wave-packet branching through an engineered conical intersection}

\author{Christopher S. Wang}
\email{christopher.wang@yale.edu}
\author{Nicholas E. Frattini}
\author{Benjamin J. Chapman}
\author{Shruti Puri}
\author{S. M. Girvin}
\author{Michel H. Devoret}
\author{Robert J. Schoelkopf}
\affiliation{Departments of Physics and Applied Physics, Yale University, New Haven, CT 06511, USA.}
\affiliation{Yale Quantum Institute, Yale University, New Haven, CT 06520, USA.}
\date{\today}

\begin{abstract}
In chemical reactions, the interplay between coherent quantum evolution and dissipation is central to determining key properties of a reaction such as the rate and yield.
Of particular interest are cases where two potential energy surfaces cross at features known as conical intersections, resulting in nonadiabatic dynamics that, under the right conditions, promote ultrafast and highly efficient reactions when rovibrational damping is present.
A prominent efficient chemical reaction that involves a conical intersection is the $\textit{cis-trans}$ isomerization reaction in rhodopsin, which is a crucial process in vision \cite{Polli2010}.
Conical intersections in real molecular systems are typically investigated via optical pump-probe spectroscopy, which has demanding spectral bandwidth and temporal resolution requirements, and where precise control of the environment is challenging.
A complementary approach for understanding chemical reactions is to use quantum simulators that can provide access to a wider range of observables than can be acquired experimentally, though thus far combining strongly interacting linear (rovibrational) and nonlinear (electronic) degrees of freedom with engineered dissipation has yet to be demonstrated.
Here, we create a tunable conical intersection in a hybrid qubit-oscillator circuit quantum electrodynamics processor and simultaneously track both a reactive wave-packet and electronic qubit in the time-domain.
We identify dephasing of the electronic qubit as the mechanism that drives wave-packet branching along the reactive coordinate in our model.
Furthermore, we directly observe enhanced branching when the wave-packet passes through the conical intersection.
Thus, the forces that influence a chemical reaction can be viewed from the perspective of measurement in quantum mechanics --- there is an effective measurement induced dephasing rate that depends on the position of the wave-packet relative to the conical intersection.
Our results set the groundwork for more complex simulations of chemical dynamics, offering deeper insight into the role of dissipation in determining macroscopic quantities of interest such as the quantum yield of a chemical reaction.
\end{abstract}

\maketitle

\begin{figure*}[t]
  \includegraphics[scale=0.95]{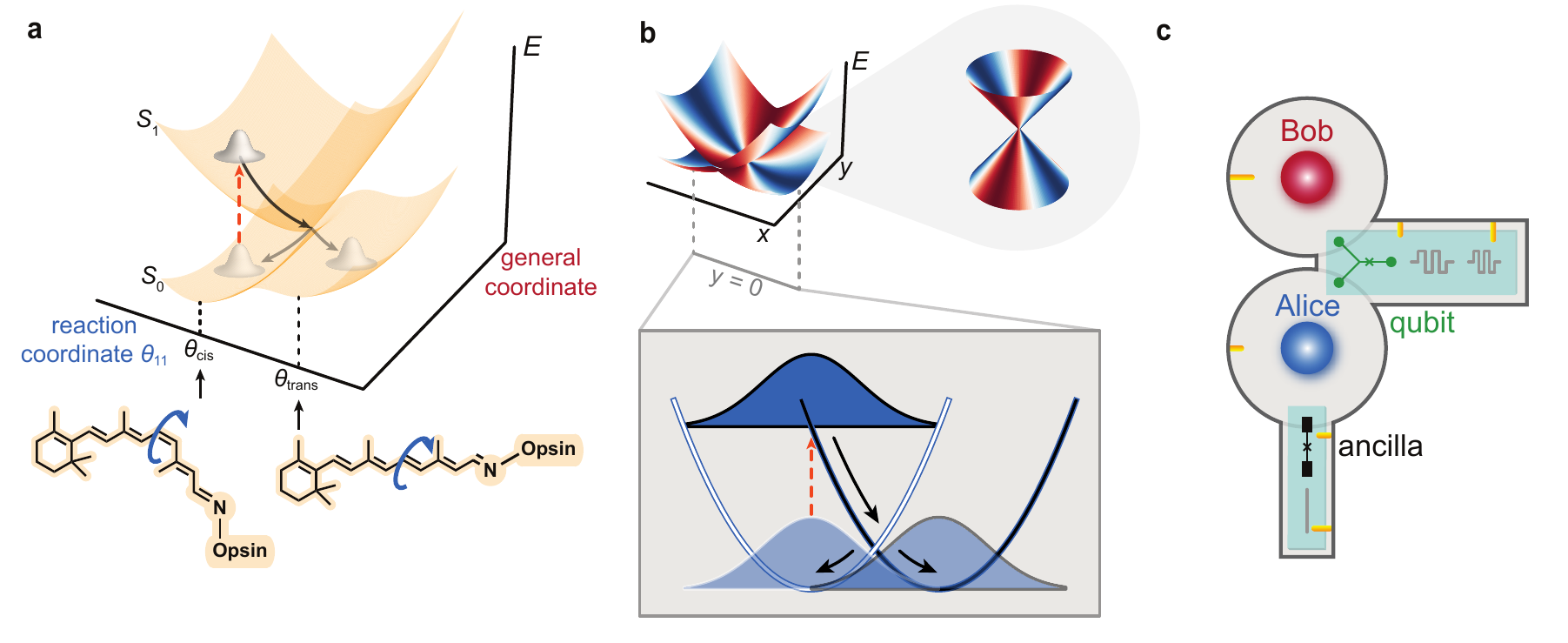}
  \caption{\textbf{Modeling a chemical reaction through a conical intersection (CI). a.} Depiction of the $\textit{cis-trans}$ isomerization reaction in rhodopsin. A ground state wave-packet becomes optically excited (dashed orange arrow) onto the upper potential energy surface $S_1$ in the Franck-Condon region. It rapidly evolves towards a CI, where it proceeds to branch either back to its reactant 11-cis configuration or towards the all-trans product. \textbf{b.} Semiclassical potential energy surfaces for an electronic qubit coupled via a CI to two harmonic confining potentials. The color depicts the quantization axis of the qubit based on the location in the $x-y$ plane (blue $\leftrightarrow \hat{\sigma}_x$ and red $\leftrightarrow \hat{\sigma}_y$). The reactive coordinate $x$ supports macroscopically distinct reactant and product ground states (linecut, bottom) associated with a two-dimensional electronic subspace $\{ \ket{\psi_1}$ (white), $\ket{\psi_2}$ (black) $\}$. \textbf{c.} Schematic of the circuit QED device used in this experiment, consisting of two 3-D $\lambda$/4 coaxial resonators, Alice (the tuning mode) and Bob (the coupling mode), coupled to a common transmon mode (the electronic degree of freedom) with a readout resonator and Purcell filter \cite{Axline2016}. An additional ancilla module with a transmon mode and readout resonator is coupled to Alice for independent state tomography. Bob is intentionally overcoupled to a 50 $\Omega$ transmission line, resulting in a linewidth $\kappa_b/2\pi \approx$ 320 kHz.}
  \label{fig:fig1}
\end{figure*}

The Born-Oppenheimer (BO) approximation is central to quantum chemistry, where the electronic and nuclear components of the molecular wavefunction are chosen to be separable owing to the typically large energy difference between electronic and nuclear motion \cite{BO1927}.
This results in adiabatic nuclear potential energy surfaces for each distinct electronic state, which are extremely valuable for visualizing and understanding nuclear wave-packet dynamics.
In many polyatomic molecules, however, two potential energy surfaces may cross at features known as conical intersections (CIs), indicating an electronic degeneracy with a local topology of a cone at a particular location in the $N - 2$ dimensional coordinate space, where $N$ is the number of rovibrational degrees of freedom.
This degeneracy invalidates the BO approximation, and the electronic and nuclear wavefunctions become strongly hybridized leading to nonadiabatic dynamics.
This hybridization has immense consequences for chemical reactions when the molecular wave-packet traverses these intersections.
One prominent example is the $\textit{cis-trans}$ isomerization reaction in rhodopsin, a chemical reaction central to vision.
There, an optically excited wave-packet rapidly evolves towards a CI and branches either back to the 11-$\textit{cis}$ reactant configuration or towards the all-trans product \cite{Polli2010}.
The quantum yield of this reaction is remarkably high ($\sim$60\%), though the exact mechanisms which enable such a high yield are not yet fully understood, in part due to computational challenges of accurately simulating such complex many-body systems coupled to structured environments \cite{Nelson2020}.

Quantum simulators may circumvent the exponentially growing computational cost of simulating larger and larger quantum systems \cite{Feynman1982}.
This cost is generally even larger when dissipation is present, particularly in regimes where bath modes may not be adiabatically eliminated.
Quantum processors applied to quantum chemistry have mostly focused on calculating electronic energies \cite{Aspuru-Guzik2005}, though experimental demonstrations have been limited to variational methods in the current near term intermediate-scale quantum (NISQ) era of devices \cite{Peruzzo2014, Kandala2017}.
Electronic transport dynamics have been investigated in the context of light harvesting in photosynthesis, where individual sites are approximated as two-level systems and encoded in qubits \cite{Potocnik2018, Maier2019}.
Vibrational dynamics and vibronic spectra have also been simulated using bosonic modes that can support multiphoton states \cite{Sparrow2018, Huh2015, Wang2020}, but only under the adiabatic BO approximation.
It is worth noting that an alternative approach for simulating bosonic rovibrational systems using a qubit-based processor is to perform a boson-to-qubit mapping \cite{Sawaya2019}, but this method incurs an overhead which will limit performance in the NISQ era.
Finally, using quantum simulators for nonadiabatic dynamics has been proposed in various platforms \cite{Ollitrault2020, MacDonell2021, Gambetta2021}, though demonstrations have remained elusive.

Here, we engineer a CI in a hybrid circuit QED processor comprising a qubit (representing the electronic excitation) and two microwave oscillators (representing generalized nuclear coordinates).
Using this, we simulate nonadiabatic dynamics of excited wave-packets and directly observe enhanced branching upon passage through the intersection.
Our approach is hardware efficient and uses a simple and optimal one-to-one mapping of the three modes of the model to three modes of the processor.
Inspired by photochemical reactions such as $\textit{cis-trans}$ isomerization in rhodopsin (Fig. 1a), we create the prototypical landscape of a chemical reaction by engineering macroscopically distinct ground states along a reactive coordinate, representing reactant and product configurations, and incorporate dissipation along a coupling coordinate.
We dynamically activate a model Hamiltonian that includes a CI using microwave drives, enabling independent characterization of the constituent interactions.
Importantly, we engineer intrinsic qubit coherence times an order of magnitude longer than the timescale of the engineered dissipation.
This corresponds to the physical scenario where a reaction is not inhibited by spontaneous emission back to the ground state in a Franck-Condon region.
By simultaneously measuring the qubit and performing Wigner tomography \cite{Bertet2002} of the state along the reactive coordinate under a control Hamiltonian, we correlate qubit dephasing events with wave-packet branching.
Finally, when the CI is active, we observe enhanced branching when the wave-packet passes through the CI, events which occur at different times depending on the initial location of the wave-packet in phase space.

\begin{figure*}[!]
  \includegraphics[scale=0.95]{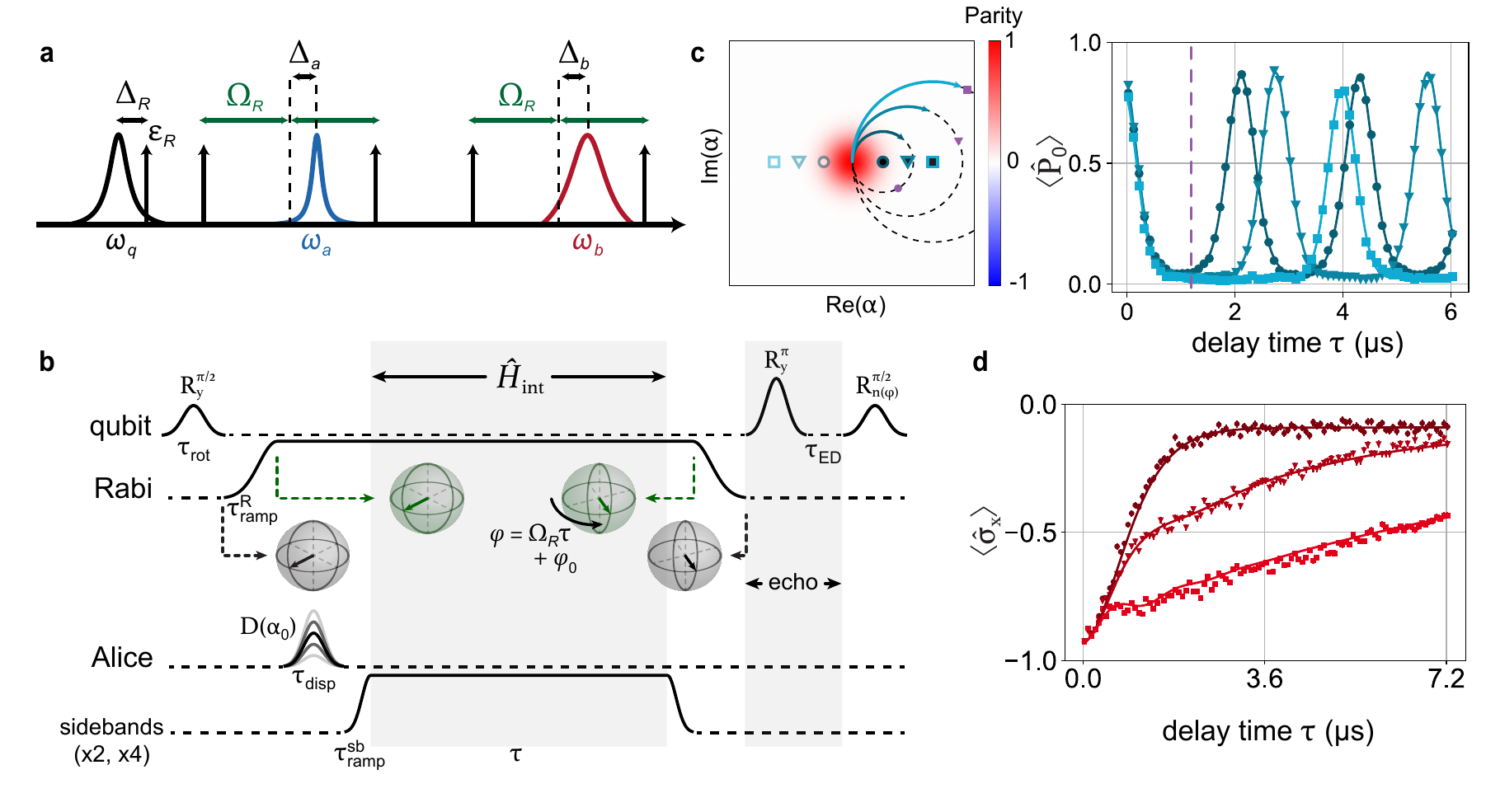}
  \caption{\textbf{Engineering conditional displacement interactions in the driven, rotating frame. a.}  Frequency configuration of the system modes and corresponding drives. The transmon is driven with a strong drive that is detuned from its g-e transition frequency, resulting in an effective energy splitting $\Omega_R$ between the two lowest driven eigenstates $\ket{\tilde{g}}$ and $\ket{\tilde{e}}$. Red- and blue-detuned sidebands separated by twice the Rabi frequency are applied to each cavity mode to enact conditional displacement interactions with the driven qubit. \textbf{b.} Pulse sequence for the full protocol (timings not to scale). The Rabi drive is ramped on slowly with respect to the static detuning $\tau^{R}_\textrm{ramp} \gg 1/\Delta_R$ to adiabatically prepare qubit eigenstates of a given conditional displacement. This finite ramp time induces residual unwanted (but deterministic) entanglement between the qubit and photons in Alice due to the cross-Kerr interaction. This necessitates both wave-packet preparation $\hat{\mathcal{D}}(\alpha_0) = e^{\alpha_0\hat{a}^\dagger - \alpha^*_0\hat{a}}$ after the Rabi drive is turned on and an echo sequence after the Rabi drive is turned off to avoid this effect (see Supplementary Information). The grey region indicates when the interaction Hamiltonian is on, where $\hat{H}_\textrm{int} \in \{\hat{H}_a, \hat{H}_b, \hat{H}_x, \hat{H}\}$ depending on which pair(s) of sidebands are turned on. Measurements are performed immediately after this sequence. \textbf{c.} Coherent state revivals revealed via measurements of the vacuum projector $\hat{P}_0 = \ket{0}\bra{0}$ on Alice (using a frequency selective $\pi$-pulse on the ancilla transmon) for an initial state $\ket{-}\otimes\ket{\alpha=0}_a$ evolving under conditional displacement interactions of different programmed values of $\Delta_a$. Solid lines are fits to a simple analytic model (see Supplementary Information), extracting values $g_x/2\pi = 450$ kHz and $\Delta_a/2\pi$ = 457 kHz (circles), 355 kHz (triangles), and 246 kHz (squares). These fit values give us values of $\alpha_g = g_x/\Delta_a$, which are depicted as the center of the circular trajectories in the cartoon on the left (not to scale). Approximate locations of the coherent state at a delay time indicated by the purple-dashed line are marked in the cartoon. \textbf{d.} Measurement-induced dephasing of $\ket{-}$ due to a conditional displacement interaction with Bob with $g_y/2\pi \approx$ 117 kHz, $\Delta_b/2\pi \approx$ $\{$ 0 (circles), 400 (triangles), and 800 (squares) $\}$ kHz and $\kappa_b/2\pi \approx$ 320 kHz. Solid lines represent time-domain master equation simulations using independently fitted parameters (Methods).}
  \label{fig:fig2}
\end{figure*}

In this work, we consider a linear vibronic coupling (LVC) model that is widely used for addressing nonadiabatic couplings between electronic and nuclear degrees of freedom \cite{Koppel1984}.
The LVC model uses a diabatic basis for the electronic eigenfunctions in order to avoid divergence issues associated with the standard adiabatic basis in the vicinity of a conical intersection \cite{Smith1969}.
Under a minimal model of two electronic states $\{ \ket{\psi_1}, \ket{\psi_2} \}$ coupled to two generalized rovibrational modes $\hat{a}$ and $\hat{b}$, we arrive at a Hamiltonian (see Supplementary Information):
\begin{equation} \label{eq:1}
\hat{H}/\hbar = \Delta_a\hat{a}^\dagger\hat{a} + \Delta_b\hat{b}^\dagger\hat{b} + g_x\hat{\sigma}_x(\hat{a} + \hat{a}^\dagger) + g_y\hat{\sigma}_y(\hat{b} + \hat{b}^\dagger)
\end{equation}
where we choose to define $\hat{\sigma}_x = \ket{\psi_1}\bra{\psi_1} - \ket{\psi_2}\bra{\psi_2}$ as the operator that represents the electronic basis (we will clarify this point later on).
Our model contains first-order intra- and inter-state couplings $g_x$ and $g_y$, respectively, as well as generalized rovibrational frequencies $\Delta_a$ and $\Delta_b$.
The modes $\hat{a}$ and $\hat{b}$ are commonly referred to as tuning and coupling modes, respectively, as the coordinate $\hat{x} \propto \hat{a} + \hat{a}^\dagger$ ``tunes" the electronic energy between $\ket{\psi_1}$ and $\ket{\psi_2}$ (Fig. 1b) and $\hat{y} \propto \hat{b} + \hat{b}^\dagger$ mediates coupling between the two electronic states via $\hat{\sigma}_y = \ket{\psi_1}\bra{\psi_2} + \ket{\psi_2}\bra{\psi_1}$.
We note that in our interest of modeling chemical reactions, we will call the position of the tuning mode the ``reactive coordinate", though we emphasize the generalized nature of these coordinates in this basis (i.e., one may need to perform a corresponding adiabatic-to-diabatic transformation).
Such a model has historical origins in the well known Jahn-Teller effect which was originally thought to necessarily be rooted in molecular symmetries, but since has been extended to larger polyatomic molecules with conical intersections that are not necessarily symmetry-induced \cite{Domcke2012}.
In general, the parameters of this model may either be empirically fit to reproduce experimental data \cite{Schneider1988} or obtained from $\textit{ab-initio}$ calculations such as for the extensively studied pyrazine \cite{Seidner1992}, the latter of which is challenging for larger polyatomic molecules.
Furthermore, the influence of various forms of rovibrational damping on conical intersection dynamics has been theoretically investigated for a number of model systems \cite{Kuhl2002, Duan2016, Schile2019}.

In this work, we operate a quantum simulator realized as a 3D circuit QED processor \cite{Wang2020} where the nuclear tuning and coupling modes are represented by $\lambda/4$ coaxial cavity modes Alice $(\hat{a})$ and Bob $(\hat{b})$, respectively, and the electronic degree of freedom is encoded in a transmon qubit \cite{Koch2007} (Fig. 1c).
The tuning mode is long-lived with a linewidth $\kappa_a/2\pi \approx$ 0.23 kHz, whereas the coupling mode is overcoupled to a 50 $\Omega$ transmission line, resulting in a decay rate $\kappa_b / 2\pi \approx 320$ kHz.
By combining a strong Rabi drive on the transmon, which results in an effective dressed qubit with an energy splitting of $\Omega_R$, with simultaneous red and blue sidebands on the cavity modes detuned by $\mp\Omega_R$ (Fig. 2a), we engineer the reaction Hamiltonian equation (\ref{eq:1}) in a driven, rotating frame.
This technique has been previously developed to simultaneously measure non-commuting qubit observables \cite{SHG2016}, suggesting an intimate link between measurement and the dissipative dynamics of our model reaction.
In contrast with previous work, our desire to control multiphoton wave-packets in the tuning mode requires us to use larger Rabi frequencies in order for the rotating wave approximation (RWA) to remain valid.
To this end, we incorporate a static detuning $\Delta_R$ on the Rabi drive and adiabatically prepare the driven qubit eigenstates to enable larger Rabi frequencies while suppressing leakage to higher levels of the transmon (see Supplementary Information).

A key requirement of our quantum simulator is the ability to initialize and perform tomography of the electronic qubit, which in our experiment is encoded by a driven transmon whose effective frequency $\Omega_R$ is defined by the amplitude $\varepsilon_R$ and static detuning $\Delta_R$ of the Rabi drive.
In the rotating frame of the drive, we define the driven qubit Hamiltonian to have the form $\hat{H}_d = \frac{\Omega_R}{2}\hat{\sigma}_z$ when expressed in the driven eigenbasis spanned by $\{ \ket{\tilde{g}}, \ket{\tilde{e}} \}$ that adiabatically connects to the undriven transmon eigenstates $\{ \ket{g}, \ket{e} \}$.
The conditional displacement interactions that we engineer in our experiment (equations \ref{eq:2},\ref{eq:3}) are conditioned on Pauli operators whose eigenstates lie on the equator of the \textit{driven} Bloch sphere (i.e., $\hat{\sigma}_x$ and $\hat{\sigma}_y$), and thus will precess around the equator at a rate $\Omega_R$.
We emphasize that our chosen convention of the Pauli operators is motivated by preferring the eigenstates of $\hat{\sigma}_z$ to be stationary in the frame of the drive.
From the perspective of the model molecular system, eigenstates of $\hat{\sigma}_x$ are the preferred electronic basis states.
We therefore use the conventional labels $\ket{\pm}$ of $\hat{\sigma}_x$ eigenstates from here onwards.
We initialize these eigenstates of $\hat{\sigma}_x$ in the driven frame by first preparing the corresponding states of the undriven transmon via a standard $\pi/2$ rotation and then adiabatically ramping on the Rabi drive $\tau^R_\textrm{ramp} \gg 1/\Delta_R$ (Fig. 2b).
In order to properly track the dynamics of any such state, we need to precisely know the Rabi frequency $\Omega_R$ so that we can decode along the appropriate axis onto our measurement basis.
By calibrating this rate, we are able to continuously measure $\langle\hat{\sigma}_{x}\rangle$ as a function of time and extract a corresponding driven coherence time $T_{2\rho}$ = 27 $\mu$s \cite{Gustavsson2012} (see Supplementary Information).

Our model Hamiltonian (equation \ref{eq:1}) consists of two simultaneous conditional displacement interactions on modes $\hat{a}$ and $\hat{b}$ with couplings to orthogonal axes of the driven qubit such that we may write $\hat{H} = \hat{H}_a + \hat{H}_b$, where
\begin{equation} \label{eq:2}
\hat{H}_a/\hbar = \Delta_a\hat{a}^\dagger\hat{a} + g_x\hat{\sigma}_x(\hat{a} + \hat{a}^\dagger)
\end{equation}
\begin{equation} \label{eq:3}
\hat{H}_b/\hbar = \Delta_b\hat{b}^\dagger\hat{b} + g_y\hat{\sigma}_y(\hat{b} + \hat{b}^\dagger).
\end{equation}
The behavior for each interaction will be qualitatively different given that we will be working in the regime where $g_x \gg \kappa_a$ for the tuning mode and $g_y \lesssim \kappa_b$ for the coupling mode.

The conditional displacement interaction on the tuning mode Alice (equation \ref{eq:2}) determines how coherent wave-packets propagate.
This Hamiltonian produces two degenerate ground states $\{ \ket{+}\otimes\ket{-\alpha_g}_a, \ket{-}\otimes\ket{+\alpha_g}_a \}$, where $\ket{\pm}$ are the eigenstates of $\hat{\sigma}_x$ and $\alpha_g = g_x/\Delta_a$.
We are interested in the regime where $\alpha_g = g_x/\Delta_a \gtrsim 1$ such that there are two macroscopically distinct ground states that can represent reactant and product.
Coherent wave-packets (which we refer to as reactive wave-packets from now on) that are prepared with the qubit in $\ket{\pm}$ at any location in phase space should oscillate around the respective ground state.
We demonstrate this first for an initial vacuum state of Alice, which is a displaced state with respect to the minimum of each of the driven potentials.
We initialize the driven qubit in $\ket{-}$ and then ramp on two sideband drives quickly with respect to $1/\Omega_R$ with the appropriate phases to enact equation (\ref{eq:2}) (see Supplementary Information).
By adjusting the average of the sideband frequencies relative to the Stark-shifted cavity frequency, we are able to tune $\Delta_a$, taking care to satisfy the resonance conditions at each frequency configuration.
This diabatically turns on the interaction and the cavity state will move along a circular trajectory in phase space around the respective ground state for the duration that the sidebands and Rabi drive are on simultaneously, which we observe by measuring the projection onto the vacuum state as a function of time using a separate transmon ancilla (Fig. 2c).
At the same time, the qubit remains in its initial eigenstate up to decoherence.
Note that if the qubit is prepared in a superposition of $\hat{\sigma}_x$ eigenstates, the conditional displacement interaction creates entanglement between the qubit and the cavity state, corresponding to a wave-packet superposition with amplitude on both potential energy surfaces.

The role of the coupling mode in our experiment, as we will see, is to induce branching of the reactive wave-packet along the potential energy surfaces.
From a quantum optics perspective, the combination of a conditional displacement interaction with single photon loss results in measurement-induced dephasing of the qubit along the axes orthogonal to the one defined by the interaction \cite{Didier2015, Touzard2019} (Methods).
We calibrate and verify this behavior by preparing an eigenstate of $\hat{\sigma}_x$, tuning the qubit axis of the interaction to be $\hat{\sigma}_y$ by adjusting the relative phase of Bob's sidebands (see Supplementary Information), and measuring the coherence as a function of time (Fig. 2d).
The detuning $\Delta_b$ is controlled in the same manner as for the interaction on Alice.
In all of our experiments, we begin with Bob near the vacuum state $\ket{0}_b$.

\begin{figure}
	\includegraphics[scale = 0.95]{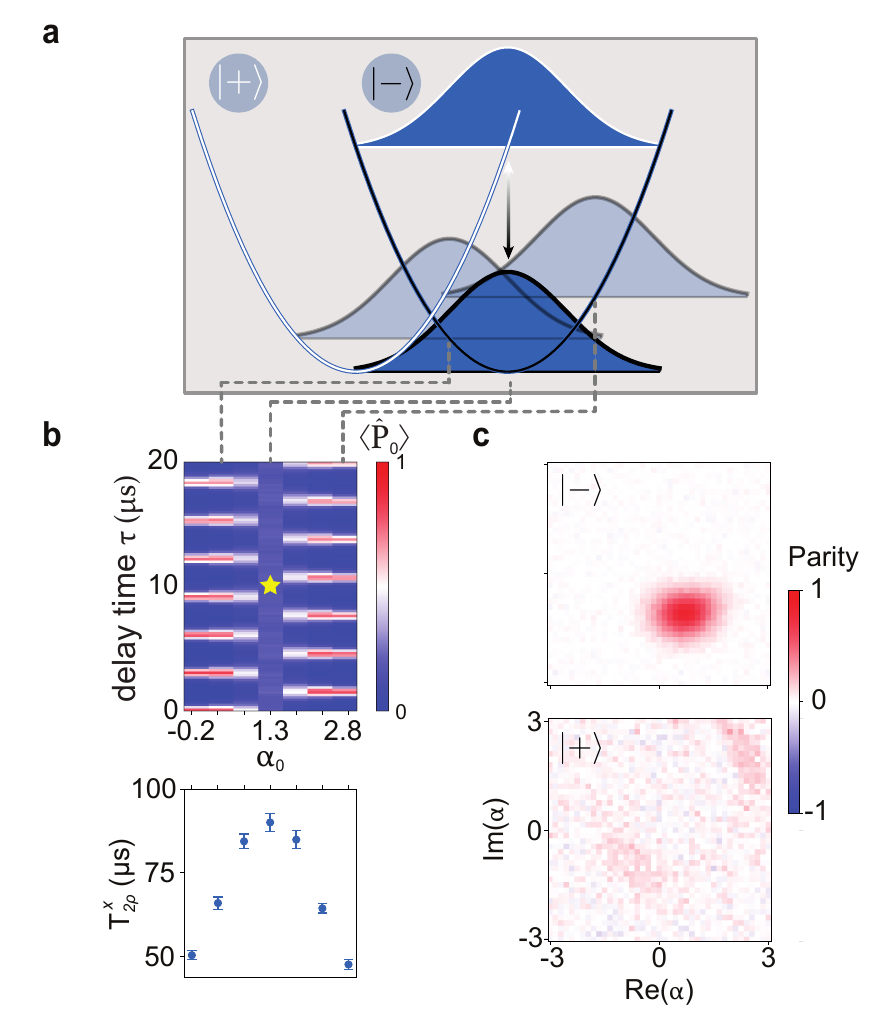}
	\caption{\textbf{Wave-packet initialization and correlated branching. a.} Depiction of various wave-packets prepared along the reactive coordinate under a conditional displacement interaction with $\hat{\sigma}_x$ of the qubit, as well as a qubit dephasing event on a ground state wave-packet. \textbf{b.} By preparing the qubit in $\ket{-}$, initializing Alice in various coherent states, and then suddenly turning on the interaction, we observe coherent revivals of the wave-packet by measuring the projection onto vacuum. The parameters here are $g_x/2\pi \approx$ 410 kHz and $\Delta_a/2\pi \approx$ 324 kHz, giving $\alpha_g = g_x/\Delta_a \approx$ 1.3. Simultaneously tracking the qubit in the $\hat{\sigma}_x$ basis reveals driven coherence times $T^x_{2\rho} > 50 \mu\textrm{s}$, with some dependence on $\langle \hat{a}^\dagger\hat{a} \rangle$. \textbf{c.} Measured Wigner functions of Alice's state conditioned on the $\hat{\sigma}_x$ measurement outcome being either $\ket{-}$ (top, $P_{-}$ = 91\%) or $\ket{+}$ (bottom, $P_{+}$ = 9\%) after evolving $\ket{\psi_0} = \ket{-}\otimes\ket{+\alpha_g}_a$ for 10 $\mu$s under the interaction (yellow star in b.). }
	\label{fig:fig3}
\end{figure}

A controllable quantum simulator must be able to perform the necessary control experiments as complexity is added.
Before enacting our model Hamiltonian, which has nontrivial and highly entangled eigenstates, we perform both conditional displacement interactions simultaneously but coupled to the same qubit axis:
\begin{equation} \label{eq:eq4}
\hat{H}_x/\hbar = \Delta_a\hat{a}^\dagger\hat{a} + \Delta_b\hat{b}^\dagger\hat{b} + g_x\hat{\sigma}_x(\hat{a} + \hat{a}^\dagger) + g_y\hat{\sigma}_x(\hat{b} + \hat{b}^\dagger)
\end{equation}
which we can engineer via aligning the phases of Bob's sidebands with respect to those on Alice.
In this scenario, the measurement-induced dephasing from the coupling mode should not perturb the dynamics of a reactive wave-packet prepared with an eigenstate of the qubit axis $\hat{\sigma}_x$.
Furthermore, we want to prepare reactive wave-packets at different locations to eventually probe the phase space dynamics of the full system (Fig. 3a).
We program our system with calibrated parameters $g_x/2\pi$ = 410 kHz, $\Delta_a/2\pi$ = 324 kHz, $g_y/2\pi \approx$ 156 kHz, and $\Delta_b/2\pi \approx$ 0 kHz.
From the above consideration, the interaction strengths for Bob's conditional displacement should not influence the dynamics of states in Alice.
We note that a photochemical reaction would involve a broadband optical excitation from a ground state to a higher potential energy surface, which in our model would correspond to performing a transition between $\hat{\sigma}_x$ eigenstates.
Alternatively, we can directly prepare the wave-packet after the optical transition.
We initialize different wave-packets by performing a displacement $\hat{\mathcal{D}}(\alpha_0)$ with the appropriate phase (see Supplementary Information) before the sidebands are activated (Fig. 2b) and observe coherent oscillations around the ground state, probed with measurements of the vacuum projector as before (Fig. 3b).

The presence of the conditional displacement interactions breaks the decoherence degeneracy between $x$ and $y$, thus we simultaneously measure the coherence time along the $x-$axis $T^x_{2\rho}$.
In our efforts to understand the decoherence that is induced on the full system solely via the coupling mode, $T^x_{2\rho}$ sets the timescale before which we need to execute our desired interactions.
Notably, we find that $T^x_{2\rho}$ is weakly dependent on the initial cavity state and peaks for the displaced ground state.
This might be explained by the presence of a residual cross-Kerr interaction, i.e. $\hat{H}_\textrm{res}/2\pi = \eta\hat{a}^\dagger\hat{a}\hat{\sigma}_z$ (see Supplementary Information), which slightly modifies the effective qubit energy in time when a non-stationary cavity state is prepared and biases the Ramsey measurement.
Furthermore, the fact that the coherence is longer compared to when the conditional displacements are off ($T_{2\rho}$) suggests an additional protection mechanism owing to the energy gap between eigenstates of $\hat{\sigma}_x$.
Nevertheless, this weak dependence does not significantly impact our results given that our characteristic dissipation-induced interaction time will be much smaller than the shortest of the measured driven coherence times.
We note, however, that explicitly controlling the relaxation of the electronic degree of freedom would be an interesting control knob in order to observe the influence of spontaneous emission on the system dynamics.

Under our model of a conditional displacement interaction, wave-packet branching between potential energy surfaces occurs when the qubit experiences a dephasing event between the electronic eigenstates, i.e. $\ket{+} \leftrightarrow \ket{-}$.
Here, qubit dephasing originates from coupling to the environment via any Pauli operator orthogonal to $\hat{\sigma}_x$.
This includes both the natural decoherence $T^x_{2\rho}$, which we investigate first here, and our eventual engineered $\hat{\sigma}_y$ coupling to the dissipative coupling mode.
A qubit dephasing event on an initial product state $\ket{\pm} \otimes \ket{\psi_\textrm{cav}}_a$ will instantaneously cause the cavity state to evolve under the opposite potential energy surface.
For example, a wave-packet that is prepared in one of the two ground states will, at random times, jump vertically to the opposite potential energy surface (Fig. 3a).
For short times compared to the coherence time $\tau < T^x_{2\rho}$, the cavity will be in a uniformly distributed mixed state (reminiscent of a donut in phase space centered around the opposite ground state) if the qubit is projected to the opposite eigenstate.
We experimentally verify this by first preparing $\ket{-}\otimes\ket{+\alpha_g}_a$ and letting the system evolve under the aligned Hamiltonian (equation (\ref{eq:eq4})) for $\tau$ = 10 $\mu$s.
Next, we measure the qubit in the $\hat{\sigma}_x$ basis and then immediately perform Wigner tomography on the cavity using the ancilla transmon \cite{Bertet2002}.
By conditioning the measured Wigner function on the electronic qubit's measurement outcome, we observe that the cavity remains in a pure state if the qubit remained in $\ket{-}$, but becomes mixed if the qubit flipped to $\ket{+}$ (Fig. 3c).
This verifies that wave-packet branching indeed occurs alongside qubit dephasing in our model.

\begin{figure}[!]
	\includegraphics[scale=0.95]{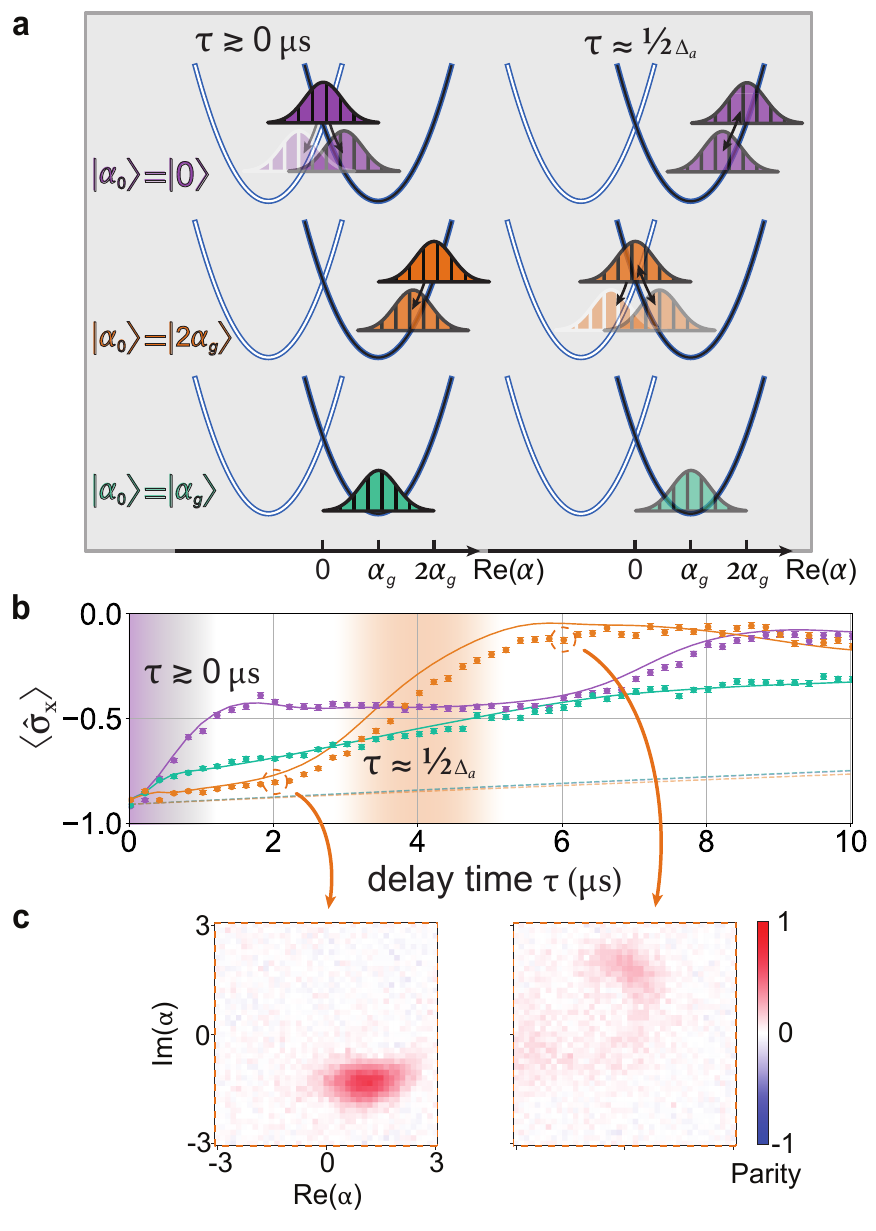}
	\caption{\textbf{Branching through a conical intersection. a.} The system is initialized with the qubit in $\ket{-}$ (black potential energy surface) with three different initial wavepackets $\ket{\alpha_0}_a \approx \{ \ket{0}_a$ (purple, top-left), $\ket{2\alpha_g}_a$ (orange, center-left), and $\ket{\alpha_g}_a$ (blue-green, bottom-left)$\}$, where $g_x/2\pi \approx$ 158.0 kHz and $\Delta_a/2\pi \approx$ 125.8 kHz giving $\alpha_g = g_x/\Delta_a \approx$ 1.26. \textbf{b.} Measured expectation value of $\hat{\sigma}_x$ over time for the three different initial states. The purple wave-packet prepared at the CI immediately dephases, whereas the other two dephase more slowly, as they are farther away. After half of an oscillation period $\tau \sim 1/(2\Delta_a)$, the orange wave-packet arrives at the CI and dephases. Solid lines are predictions from a master equation simulation using independently fitted parameters. Dashed lines represent the negligible background decoherence due to $T^x_{2\rho}$ on the timescale of the interaction and dissipation. \textbf{c.} Unconditional Wigner tomography on Alice at $\tau$ = 2 $\mu$s (left) and 6 $\mu$s (right) for preparing $\ket{\alpha_0}_a \approx \ket{2\alpha_g}_a$, revealing a coherent wave-packet before and dephased state after passage through the CI. The distortion of the Wigner function from a Gaussian at $\tau$ = 2 $\mu$s suggests the presence of a residual self-Kerr nonlinearity in the oscillator.}
	\label{fig:fig4}
\end{figure}

The primary task of our experiment is to understand the time dynamics of excited reactive wave-packets, particularly through the CI.
With our understanding that qubit dephasing along $\hat{\sigma}_x$ drives wave-packet branching, we prepare the qubit in $\ket{-}$ and directly monitor $\langle\hat{\sigma}_x\rangle$ as a function of time for different initial wave-packet configurations in the tuning mode (Fig. 4a).
Here, we program our system with calibrated parameters $g_x/2\pi$ = 158.0 kHz, $\Delta_a/2\pi$ = 125.8 kHz, $g_y/2\pi$ = 115 kHz, and $\Delta_b/2\pi \approx$ 0 kHz.
In this instance, we reduce the interaction strengths of the reactive potential surface to get a clear signature of branching over the course of one period of motion, and choose zero detuning on the coupling mode's conditional displacement to achieve the strongest dephasing, representing a very slow intranuclear mode.
We indeed observe the hallmark of dynamics through the CI - the qubit dephasing is both highly non-exponential and stronger upon passage of the wave-packet through the intersection (Fig. 4b).
Additionally, we further verify that this behavior indeed correlates with dephasing of the cavity state by taking Wigner functions of an initially displaced wave-packet before and after passage through the CI (Fig. 4c).
In our case, preparation of an excited wave-packet in the Franck-Condon region (Fig. 1b) leads to large photon numbers that enact higher order Rabi frequency shifts that bias our measurement.

We can qualitatively understand the decoherence behavior by treating the tuning mode classically.
Here, we are left with a simplified Hamiltonian
\begin{equation}
\hat{H}_\textrm{Zeno}/\hbar = E(x)\hat{\sigma}_x + \Delta_b\hat{b}^\dagger\hat{b} + g_y\hat{\sigma}_y(\hat{b} + \hat{b}^\dagger)
\end{equation}
subject to single photon loss on mode $\hat{b}$ at a rate $\kappa_b$. 
The function $E(x)$ can be interpreted as the position dependent energy gap (in frequency units) between qubit states for a conditional displacement interaction.
At the CI, i.e., at $x = 0$, this energy gap vanishes and we are left with the environment measuring $\hat{\sigma}_y$ with a measurement strength $\Gamma_\textrm{meas} = g^2_y\kappa_b/[(\kappa_b/2)^2 + \Delta^2_b]$ in the steady state where $g_y \ll \kappa_b$.
Away from the origin, the qubit has a finite energy along an orthogonal axis to that of the measurement and we recover a scenario reminiscent of Zeno dynamics of a driven qubit \cite{Gambetta2008}.
This reduces the effective measurement strength, resulting in slower decoherence and thus reduced branching events (see Supplementary Information).
Our experiment may qualitatively be understood from this perspective by choosing a time-dependent trajectory $x(t)$ for an initial Gaussian wave-packet.
In our full model, wave-packets in the tuning mode will diffuse in phase space due to the branching, resulting in dynamics that are quantitatively different from the above simplified model.

A number of modifications to our experimental setup can be made to expand the scope of our reaction model.
For instance, by overcoupling Alice to a transmission line, the reactive coordinate will experience dissipation which would eventually localize an initially excited wave-packet into the two ground states.
The addition of $\kappa_a$ as a tuning knob, particularly in the regime $\kappa_a\sim\kappa_b$, expands the landscape of competing forces in our model and represents a more realistic description of a reaction by defining a quantum yield (see Supplementary Information).
We note the possibility of tuning the decay rate of each oscillator $\textit{in-situ}$ via mechanical means in our 3D architecture, which would enable a flexible way to explore the wider range of parameter space.
Furthermore, expanding our simulator's capabilities to incorporate anharmonicities in the potential energy surfaces allows for a more accurate modeling of realistic systems, whose dynamics are heavily influenced by nuclear topography.
These systems may also have the Franck-Condon region far from the CI, which will translate to larger photon numbers in our system.
As such, precisely controlling the desired nonlinear reaction Hamiltonian over the domain of larger photon numbers will be one challenge to address in future experiments.
Finally, the finite anharmonicity of the transmon sets a limit on how large the Rabi frequency can be while preserving the qubit nature of the driven interaction.
This constrains how strong the interactions can be while respecting the RWA, motivating the investigation of alternate qubit modalities.

Our results highlight the interplay between coherent evolution and engineered dissipation in a system whose energy landscape contains a CI.
We achieve this via careful Hamiltonian engineering of a circuit QED processor involving five simultaneous microwave drives and engineered dissipation, along with the appropriate state preparation and measurement protocols to observe branching dynamics.
In particular, we identify branching events to arise when dissipation in the coupling mode induces flips on the electronic state and correspondingly causes the reactive wave-packet to jump onto the opposite potential surface.
These branching events are at the heart of chemical reaction dynamics, and occur more frequently upon passage through the CI.
Our experiment demonstrates and further confirms the immense flexibility of this platform to perform quantum information processing tasks by dressing microwave photons with continuous drives \cite{SHG2016, Grimm2020, Burkhart2021}, and constitutes an important step towards investigating more complex chemical phenomena with higher degrees of accuracy.
Furthermore, we also identify the boundaries of the approach taken in this experiment and how they inform the design of next generation devices and control schemes.
It is worth noting that the techniques developed in our work may readily be applicable to control multi-mode bosonic systems coupled to one or a few qubits \cite{Chakram2020, Owens2021}.
Looking ahead, incorporating additional features to our simulator such as additional controlled nonlinearities and structured dissipators unlock new regimes that may provide deeper insight into chemical phenomena.
Broadly, this expands the landscape of controllable qubit-oscillator interactions in a circuit QED platform, which may be useful for bosonic quantum computation, error correction, and simulation.

\begin{acknowledgments}
We acknowledge the early insight of V. Batista for simulating conical intersections. We thank A. Eickbusch, R. Cortiñas, and A. Koottandavida for helpful discussions in guiding experimental details. We acknowledge Y. Gao and B. Lester for initial package design. Facilities use was supported by YINQE and the Yale SEAS cleanroom. This research is supported by the Army Research Office (ARO) under Grant No. W911NF-18-1-0212. The views and conclusions contained in this document are those of the authors and should not be interpreted as representing the official policies, either expressed or implied, of the ARO, or the U.S. Government. The U.S. Government is authorized to reproduce and distribute reprints for Government purposes notwithstanding any copyright notation herein. S.P. acknowledges support from the National Science Foundation (QLCI grant OMA-2120757). We also acknowledge support of the Yale Quantum Institute.
\end{acknowledgments}

%

\section*{Methods}

\section*{Experimental Details}
The cQED device used in this work closely resembles those used in Refs. \cite{Gao2019, Wang2020}. 
The primary difference is that in those experiments, the coupling transmon was optimized to perform as a nonlinear mixing element for parametric processes between the two cavities and thus had a relatively small anharmonicity $\alpha_q/2\pi \approx$ 70 MHz.
Here, we instead operate the coupling transmon in a more standard qubit-like regime $\alpha_q/2\pi \approx$ 240 MHz which enables us to enact a strong Rabi drive and still primarily operate in the manifold of the two lowest energy eigenstates.
A full table of system parameters is given in the Supplementary Information.

\section*{Enacting the Model Hamiltonian}
We enact our model Hamiltonian (equation (\ref{eq:1}) of the main text) using a technique that is similar to and inspired by Ref. \cite{SHG2016}, which involves combining a Rabi drive on the qubit with simultaneous red and blue detuned sidebands on each of the cavity modes.
We modify their protocol slightly in order to fully control and measure our system by adding a static detuning parameter $\Delta_R$ on the Rabi drive, which translates to adiabatic preparation of the driven eigenstates.
This modification enables higher fidelity control of the cavity states by improving the degree in which the RWA is valid for the interaction Hamiltonian by allowing higher Rabi frequencies $\Omega_R$ while simultaneously suppressing leakage to higher excited states of the transmon.
It also is crucial for performing qubit tomography to cancel the residual cross-Kerr between the driven Rabi qubit and the cavity modes.
An undesired consequence of this modification is the presence of a residual entanglement interaction between the qubit and photons in the cavities during the finite ramp time of the Rabi drive, but we undo this deterministic effect by performing a short echo sequence (Fig. 2b).
These details are fully expanded on in the Supplementary Information.

\section*{Master Equation Simulations}
The theoretical predictions for $\langle\hat{\sigma}_x\rangle$ in Figs. 2d and 4b are obtained by performing a numerical simulation of a full time-domain master equation using a Python-based open source software (QuTiP):
\begin{equation*}
\dot{\hat{\rho}} = -\frac{i}{\hbar}[\hat{H}_\textrm{sim},\hat{\rho}] + \kappa_b\mathcal{D}[\hat{b}]\hat{\rho} + \frac{\gamma_y}{2}\mathcal{D}[\hat{\sigma}_y]\hat{\rho}
\end{equation*}
where the left hand side of the equation is the time derivative of the system's density matrix and $\mathcal{D}[\hat{A}]\hat{\rho} = \hat{A}\hat{\rho}\hat{A}^\dagger - \frac{1}{2}\hat{A}^\dagger\hat{A}\hat{\rho} - \frac{1}{2}\hat{\rho}\hat{A}^\dagger\hat{A}$.
We extract $\langle\hat{\sigma}_x\rangle$ by plotting Tr[$\hat{\rho}^\textrm{qubit}\hat{\sigma}_x$], where $\hat{\rho}^\textrm{qubit}$ = Tr$_\textrm{a,b}\hat{\rho}$ is the reduced density matrix of the qubit only.
For the measurement-induced dephasing in Fig. 2d, we use $\hat{H}_\textrm{sim}$ = $\hat{H}_b$ from equation (\ref{eq:3}) and an initial state $\ket{-}\otimes\ket{0}_b$ for various values of $\Delta_y$.
For the full system dynamics through the conical intersection in Fig. 4b, we use $\hat{H}_\textrm{sim} = \hat{H}$ from equation (\ref{eq:1}) and initial states $\ket{-}\otimes\ket{\alpha_0}_a\otimes\ket{0}_b$ for various values of $\alpha_0$.

Across these two simulations, the parameters $\{ g_x, \Delta_a, g_y, \kappa_b \}$ are determined via independent calibrations, and $\Delta_b$ is estimated via the frequencies of Bob's sidebands.
The value of $\gamma_y$ is set by an independent measurement of $T^x_{2\rho}$, specifically via $\gamma_y = 1/T^x_{2\rho}$.
Finally, the amplitude and offset are adjusted using the measured experimental values from the corresponding $T^x_{2\rho}$ control measurement, leaving zero free parameters.



\clearpage
\begin{thebibliography}{9}

\bibitem{Polli2010}
D. Polli, P. Alto\'{e}, O. Weingart, K. M. Spillane, C. Manzoni, D. Brida, G. Tomasello, G. Orlandi, P. Kukura, R.A. Mathies, M. Garavelli, and G. Cerullo, Conical intersection dynamics of the primary photoisomerization event in vision, \textit{Nature} 467, 440 (2010).

\bibitem{BO1927}
M. Born and R. Oppenheimer. On the Quantum Theory of Molecules, \textit{Ann. Phys.} 84, 457 (1927).

\bibitem{Nelson2020}
T. Nelson, A. White, J. Bjorgaard, A. Sifain, Y. Zhang, B. Nebgen, S. Fernandez-Alberti, D. Mozyrsky, A. Roitberg, and S. Tretiak, Non-adiabatic Excited-State Molecular Dynamics: Theory and Applications for Modeling Photophysics in Extended Molecular Materials, \textit{Chem. Rev.} 120, 2215 (2020).

\bibitem{Feynman1982}
R. P. Feynman, Simulating physics with computers, \textit{Int. J. Theor. Phys.} 21, 467–488 (1982).

\bibitem{Aspuru-Guzik2005}
A. Aspuru-Guzik, A. D. Dutoi, P. J. Love, and M. Head- Gordon, Simulated Quantum Computation of Molecular Energies, \textit{Science} 309, 1704 (2005).

\bibitem{Peruzzo2014}
A. Peruzzo, J. McClean, P. Shadbolt, M.-H. Yung, X.-Q. Zhou, P. J. Love, A. Aspuru-Guzik, and J. L. O’Brien, A Variational Eigenvalue Solver on a Photonic Quantum Processor, \textit{Nature Communications} 5, 1 (2014).

\bibitem{Kandala2017}
A. Kandala, A. Mezzacapo, K. Temme, M. Takita, M. Brink, J. M. Chow, and J. M. Gambetta, Hardware- Efficient Variational Quantum Eigensolver for Small Molecules and Quantum Magnets, \textit{Nature} 549, 242 (2017).

\bibitem{Potocnik2018}
A. Potocnik, A. Bargerbos, F. A.Y.N. Schroder, S. A. Khan, M. C. Collodo, S. Gasparinetti, Y. Salathe, C. Creatore, C. Eichler, H. E. Tureci, A. W. Chin, and A. Wallraff, Studying Light-Harvesting Models with Superconducting Circuits, \textit{Nature Communications.} 9, 904 (2018).

\bibitem{Maier2019}
C. Maier, T. Brydges, P. Jurcevic, N. Trautmann, C. Hempel, B. P. Lanyon, P. Hauke, R. Blatt, and C. F. Roos, \textit{Phys. Rev. Lett.} 122, 050501 (2019).

\bibitem{Sparrow2018}
C. Sparrow, E. Martin-Lopez, N. Maraviglia, A. Neville, C. Harrold, J. Carolan, Y. N. Joglekar, T. Hashimoto, N. Matsuda, J. L. O’Brien, D. P. Tew, and A. Laing, Simulating the vibrational quantum dynamics of molecules using photonics, \textit{Nature.} 557, 660 (2018).

\bibitem{Huh2015}
J. Huh, G. G. Guerreschi, B. Peropadre, J. R. McClean, and A. Aspuru-Guzik, Boson sampling for molecular vibronic spectra, \textit{Nature Photonics.} 9, 615–620 (2015).

\bibitem{Wang2020}
C. S. Wang, J. C. Curtis, B. J. Lester, Y. Zhang, Y. Y. Gao, J. Freeze, V. S. Batista, P. H. Vaccaro, I. L. Chuang, L. Frunzio, L. Jiang, S. M. Girvin, and R. J. Schoelkopf, Efficient Multiphoton Sampling of Molecular Vibronic Spectra on a Superconducting Bosonic Processor, \textit{Phys. Rev. X} 10, 021060 (2020).

\bibitem{Sawaya2019}
N. P. D. Sawaya and J. Huh, Quantum Algorithm for Calculating Molecular Vibronic Spectra, \textit{J. Phys. Chem. Lett.} 10, 3586 (2019).

\bibitem{Ollitrault2020}
P. J. Ollitrault, G. Mazzola, and I. Tavernelli, Nonadiabatic Molecular Quantum Dynamics with Quantum Computers, \textit{Phys. Rev. Lett.} 125, 260511 (2020).

\bibitem{MacDonell2021}
R. J. MacDonell, C. E. Dickerson, C. J. T. Birch, A. Kumar, C. L. Edmunds, M. J. Biercuk, C. Hempel, and I. Kassal, Analog quantum simulation of chemical dynamics, \textit{Chem. Sci.} 12, 9794-9805 (2021).

\bibitem{Gambetta2021}
F. M. Gambetta, C. Zhang, M. Hennrich, I. Lesanovsky, and W. Li, Exploring the Many-Body Dynamics Near a Conical Intersection with Trapped Rydberg Ions, \textit{Phys. Rev. Lett.} 126, 233404 (2021).

\bibitem{Bertet2002}
P. Bertet, A. Auffeves, P. Maioli, S. Osnaghi, T. Meunier, M. Brune, J. M. Raimond, and S. Haroche, Direct Meas- urement of the Wigner Function of a One-Photon Fock State in a Cavity, \textit{Phys. Rev. Lett.} 89, 200402 (2002).


\bibitem{Koppel1984}
H. Köppel, W. Domcke and L. S. Cederbaum, Multimode molecular dynamics beyond the Born-Oppenheimer approximation, \textit{Adv. Chem. Phys.} 57, 59 (1984).

\bibitem{Smith1969}
F. T. Smith, Diabatic and Adiabatic Representations for Atomic Collision Problems, \textit{Phys. Rev.} 179, 111 (1969).

\bibitem{Domcke2012}
W. Domcke and D. R. Yarkony, Role of Conical Intersections in Molecular Spectroscopy and Photoinduced Chemical Dynamics, \textit{Annu. Rev. Phys. Chem.} 63, 325 (2012).

\bibitem{Schneider1988}
R. Schneider and W. Domcke, S1-S2 Conical intersection and ultrafast S2 $\rightarrow$ S1 internal conversion in pyrazine, \textit{Chem. Phys. Lett.} 150, 235 (1988).

\bibitem{Seidner1992}
L. Seidner, G. Stock, A. Sobolewski, and W. Domcke, Ab initio characterization of the S1–S2 conical intersection in pyrazine and calculation of spectra, \textit{J. Chem. Phys.} 96, 5298 (1992).

\bibitem{Kuhl2002}
A. Kühl and W. Domcke, Multilevel Redfield description of the dissipative dynamics at conical intersections, \textit{J. Chem. Phys.} 116, 263 (2002).

\bibitem{Duan2016}
H.-G. Duan and M. Thorwart, Quantum Mechanical Wave Packet Dynamics at a Conical Intersection with Strong Vibrational Dissipation, \textit{J. Phys. Chem. Lett.} 7, 382 (2016).

\bibitem{Schile2019}
A. J. Schile and D. T. Limmer, Simulating Conical Inter- section Dynamics in the Condensed Phase with Hybrid Quantum Master Equations, \textit{J. Chem. Phys.} 151, 014106 (2019).

\bibitem{Koch2007}
J. Koch, T. M. Yu, J. Gambetta, A. A. Houck, D. I. Schuster, J. Majer, A. Blais, M. H. Devoret, S. M. Girvin, R. J. Schoelkopf, Charge-insensitive qubit design derived from the Cooper pair box, \textit{Phys. Rev. A.} 76, 042319 (2007).

\bibitem{Axline2016}
C. Axline, M. Reagor, R. Heeres, P. Reinhold, C. Wang, K. Shain, W. Pfaff, Y. Chu, L. Frunzio, and R. J. Schoelkopf, An Architecture for Integrating Planar and 3D cQED Devices, \textit{Appl. Phys. Lett.} 109, 042601 (2016).

\bibitem{SHG2016}
S. Hacohen-Gourgy, L. S. Martin, E. Flurin, V. V. Ramasesh, K.B. Whaley, and I. Siddiqi, Quantum Dynamics of Simultaneously Measured Non-Commuting Observables, \textit{Nature} 538, 491 (2016).

\bibitem{Gustavsson2012}
S. Gustavsson et al., Driven Dynamics and Rotary Echo of a Qubit Tunably Coupled to a Harmonic Oscillator, \textit{Phys. Rev. Lett.} 108, 170503 (2012).

\bibitem{Didier2015}
N. Didier, J. Bourassa, and A. Blais, Fast Quantum Nondemolition Readout by Parametric Modula- tion of Longitudinal Qubit-Oscillator Interaction, \textit{Phys. Rev. Lett.} 115, 203601 (2015).

\bibitem{Touzard2019}
S. Touzard, A. Kou, N. E. Frattini, V. V. Sivak, S. Puri, A. Grimm, L. Frunzio, S. Shankar, and M. H. Devoret, Gated Conditional Displacement Readout of Superconducting Qubits, \textit{Phys. Rev. Lett.} 122, 080502 (2019).

\bibitem{Gambetta2008}
J. Gambetta, A. Blais, M. Boissonneault, A.A. Houck, D. I. Schuster, and S. M. Girvin, Quantum trajectory approach to circuit QED: Quantum jumps and the Zeno effect, \textit{Phys. Rev. A} 77, 012112 (2008).

\bibitem{Grimm2020}
A. Grimm, N. E. Frattini, S. Puri, S. O. Mundhada, S. Touzard, M. Mirrahimi, S. M. Girvin, S. Shankar, and M. H. Devoret, Stabilization and operation of a Kerr-cat qubit, \textit{Nature} 584, 205 (2020).

\bibitem{Burkhart2021}
L. D. Burkhart, J. Teoh, Y. Zhang, C. J. Axline, L. Frunzio, M. H. Devoret, L. Jiang, S. M. Girvin, and R. J. Schoelkopf, Error-detected state transfer and entanglement in a superconducting quantum network, \textit{PRX Quantum.} 2, 030321 (2021).

\bibitem{Chakram2020}
S. Chakram, K. He, A. V. Dixit, A. E. Oriani, R. K. Naik, N. Leung, H. Kwon, W.-L. Ma, L. Jiang, and D. I. Schuster, Multimode photon blockade, \textit{arXiv:2010.15292} (2020).

\bibitem{Owens2021}
J. C. Owens, M. G. Panetta, B. Saxberg, G. Roberts, S. Chakram, R. Ma, A. Vrajitoarea, J. Simon, D. Schuster, Chiral Cavity Quantum Electrodynamics, \textit{arXiv:2109.06033} (2021).

\bibitem{Gao2019}
Y. Y. Gao, B. J. Lester, K. S. Chou, L. Frunzio, M. H. Devoret, L. Jiang, S. M. Girvin, and R. J. Schoelkopf, Entanglement of bosonic modes through an engineered exchange interaction, \textit{Nature} 566, 509 (2019).

\end{thebibliography}
\end{document}


\preprint{}

\title{Supplemental Material: Observation of wave-packet branching through an engineered conical intersection}

\author{Christopher S. Wang}
\email{christopher.wang@yale.edu}
\author{Nicholas E. Frattini}
\author{Benjamin J. Chapman}
\author{Shruti Puri}
\author{S. M. Girvin}
\author{Michel H. Devoret}
\author{Robert J. Schoelkopf}
\affiliation{Departments of Physics and Applied Physics, Yale University, New Haven, CT 06511, USA.}
\affiliation{Yale Quantum Institute, Yale University, New Haven, CT 06520, USA.}
\date{\today}

\maketitle

\section{Constructing a model reaction Hamiltonian containing a conical intersection}

In this section, we describe the general formalisms behind obtaining various model and/or $\textit{ab-initio}$ molecular Hamiltonians that involve strong vibronic coupling.
We begin with a brief review of adiabatic potential energy surfaces and highlight the difficulties that arise in the vicinity of conical intersections.
This motivates the use of diabatic electronic states, which form the basis of the Hamiltonians that we consider \cite{Domcke2004}.

The standard molecular Hamiltonian is:
\be
\hat{H}_\textrm{mol} = \hat{T}_n(\partial_{\vec{R}}) + \underbrace{\hat{T}_e(\partial_{\vec{r}}) + \hat{V}(\vec{r}, \vec{R})}_{\hat{H}_e}
\ee
where $\hat{T}$ and $\hat{V}$ correspond to kinetic and Coulomb potential energies, respectively, and the subscripts $n$ and $e$ denote nuclei and electrons. $\vec{r}$ and $\vec{R}$ represents the positions of the electrons and nuclei, respectively.

The conventional approach based on the Born-Oppenheimer approximation begins with noting that the electron masses $m_e$ are much smaller than that of the nuclear masses $m_N$.
This motivates momentarily dropping $\hat{T}_n(\partial_{\vec{R}}) \propto 1/m_N$ altogether, leaving behind a reduced Hamiltonian of the electrons only with parametric dependence on the nuclear coordinates $\vec{R}$.
By choosing an ansatz for the total molecular wave-function $\Psi(\vec{r},\vec{R}) = \Sigma_i\varphi_i(\vec{r},\vec{R})\chi_i(\vec{R})$, we get a reduced electronic Schrodinger equation:
\be
\hat{H}_e\varphi_i(\vec{r},\vec{R}) = E_i(\vec{R})\varphi_i(\vec{r},\vec{R})
\ee
where $E_i(\vec{R})$ is the potential energy surface for the $i^{th}$ electronic eigenstate.
Here, we have identified a complete set of adiabatic electronic eigenfunctions $\varphi_i(\vec{r},\vec{R})$.
Returning to the full Schrodinger equation
\be
\hat{H}_\textrm{mol}\Psi(\vec{r},\vec{R}) = (\hat{T}_n(\partial_{\vec{R}}) + \hat{H}_e)\sum_i\varphi_i(\vec{r},\vec{R})\chi_i(\vec{R}),
\ee
we can obtain a reduced equation for the nuclear motion by integrating over a complete set of adiabatic electronic eigenfunctions $\varphi^*_j(\vec{r},\vec{R})$.
This brings out terms such as:
\be
\frac{\bra{\varphi_j(\vec{r},\vec{R})}\hat{\nabla}_R\hat{H}_e\ket{\varphi_i(\vec{r},\vec{R})}}{E_j(\vec{R}) - E_i(\vec{R})}
\ee
which arise from the application of $\hat{T}_n(\partial_{\vec{R}}) \propto -\frac{1}{2m_N}\hat{\nabla}^2_R$ on $\varphi_i(\vec{r},\vec{R})$.
These are commonly referred to as non-adiabatic coupling terms (NACT) in the literature.
It is clear that in the vicinity of a conical intersection, these terms diverge as the denominator becomes very small and the adiabatic electronic basis fails to be an appropriate basis for calculations and analyses.

Given the aforementioned issue, one can consider instead a diabatic electronic basis $\phi_k(\vec{r})$ such that the molecular wave-function can be expressed as:
\be
\Psi(\vec{r},\vec{R}) = \sum_k \phi_k(\vec{r})\chi'_k(\vec{R})
\ee
where the diabatic states are, by definition, diagonal in the nuclear kinetic energy operator.
Off-diagonal couplings between electronic states must of course exist, but now they arise via the potential $\bra{\phi_j}\hat{V}(\vec{r},\vec{R})\ket{\phi_i}$ and do not involve wave-function derivatives.

This forms the basis for a general vibronic coupling Hamiltonian:
\be
\hat{H}_\textrm{vc} = \sum_n\ket{\phi_n}(\hat{T}_n + W_{nn}(\vec{R}))\bra{\phi_n} + \sum_{n \neq m} \ket{\phi_n}W_{nm}(\vec{R})\bra{\phi_m}
\ee
In our experiment, we consider a minimal model for a two-dimensional linear vibronic coupling (LVC) Hamiltonian, where we have:
\begin{align}
\hat{T}_n & = \frac{\hat{p}^2_x}{2m_x} + \frac{\hat{p}^2_y}{2m_y} \\
\hat{W} & = \left(\begin{array}{cc} \frac{1}{2}\omega^2_x \hat{x}^2 + \frac{1}{2}\omega^2_y \hat{y}^2 + g_x \hat{x} & g_y \hat{y} \\ g_y \hat{y} & \frac{1}{2}\omega^2_x \hat{x}^2 + \frac{1}{2}\omega^2_y \hat{y}^2 - g_x \hat{x} \end{array}\right)
\end{align}
for two normal coordinates $\hat{x}$ and $\hat{y}$.
By re-casting these coordinates into creation and annihilation operators $\hat{x} \propto (\hat{a} + \hat{a}^\dagger), \hat{y} \propto (\hat{b} + \hat{b}^\dagger)$, we arrive at Eq. (1) of the main text.

\section{Engineering the conical intersection interaction}

Here, we show how we engineer the reaction Hamiltonian (Eq. 1 of the main text).
As discussed in the main text, this Hamiltonian consists of two simultaneous conditional displacement interactions of a single qubit to two different cavity modes, where the qubit coupling axes are orthogonal.
To simplify the derivation, we will begin by focusing on how we enact just one of these interactions.
As we will see, adding additional interactions to other cavity modes is relatively straightforward, and the qubit coupling axis is freely adjustable in the effective $x-y$ plane.

We expand upon the derivation provided by \cite{SHG2016} by incorporating the finite anharmonicity of the transmon mode.
We begin with the static Hamiltonian of a transmon mode $\hat{q}$ dispersively coupled to a cavity mode $\hat{c}$:
\be
\hat{H}_\textrm{static}/\hbar = \omega_c\hat{c}^\dagger\hat{c} + \omega_q\hat{q}^\dagger\hat{q} -\frac{\alpha_q}{2}\hat{q}^\dagger\hat{q}^\dagger\hat{q}\hat{q} -\chi\hat{c}^\dagger\hat{c}\hat{q}^\dagger\hat{q}
\ee
where $\alpha_q$ is the transmon anharmonicity and $\chi$ is the dispersive shift.
At a high level, we will see that the conditional displacement interaction arises by transforming the cross-Kerr interaction between the transmon and the cavity.
Thus, our approach will be to consider how driving each mode transforms the static interaction.
Specifically, we drive the system with one tone coupled to the transmon and two coupled to the cavity:
\be
\hat{H}_d/\hbar = 2\varepsilon_R\textrm{cos}[(\omega_q + \Delta_R)t](\hat{q} + \hat{q}^\dagger) -2i\varepsilon_1\textrm{sin}(\omega_1t + \varphi_1)(\hat{c} - \hat{c}^\dagger) -2i\varepsilon_2\textrm{sin}(\omega_2t + \varphi_2)(\hat{c} - \hat{c}^\dagger) 
\ee
such that the full system Hamiltonian is described by $\hat{H} = \hat{H}_\textrm{static} + \hat{H}_d$.
For convenience, we re-group the terms such that we can write $\hat{H} = \hat{H}_q(\hat{q}, \hat{q}^\dagger) + \hat{H}_c(\hat{c}, \hat{c}^\dagger) + \hat{H}_\textrm{int}$, where $\hat{H}_\textrm{int}/\hbar = -\chi\hat{c}^\dagger\hat{c}\hat{q}^\dagger\hat{q}$.
We first go into the rotating frame of the transmon drive via $\hat{H} \rightarrow \hat{U}\hat{H}\hat{U}^\dagger + i\dot{\hat{U}}\hat{U}^\dagger$, where $\hat{U} = e^{i(\omega_q + \Delta_R)t\hat{q}^\dagger\hat{q}}$:
\be
\hat{H}/\hbar = \underbrace{- \Delta_R\hat{q}^\dagger\hat{q} -\frac{\alpha_q}{2}\hat{q}^\dagger\hat{q}^\dagger\hat{q}\hat{q} + \varepsilon_R(\hat{q} + \hat{q}^\dagger)}_{\hat{H}_\textrm{q}/\hbar} + \hat{H}_c/\hbar -\chi\hat{c}^\dagger\hat{c}\hat{q}^\dagger\hat{q}
\ee
noting that we have performed the rotating-wave approximation (RWA) and discarded terms rotating at $\mathcal{O}(\omega_q)$. Furthermore, the cross-Kerr term remains unaffected since it is proportional to $\hat{q}^\dagger\hat{q}$.
Now, we diagonalize $\hat{H}_q$ and re-express it in the resulting eigenbasis:
\be
\hat{H}_\textrm{q}/\hbar = \sum_i \epsilon_i \ket{i}\bra{i}
\ee
where we label $i \in \{ +, -, \tilde{f}, ...\}$ in correspondence with the fact that we will be working in a regime where the two lowest driven eigenstates strongly resemble those of a standard qubit that is driven on resonance, but now incorporate a weak dressing with higher levels of the transmon.
We identify the Rabi frequency to be the energy difference between the lowest two eigenstates $\epsilon_+ - \epsilon_- = \Omega_R$ and define an effective anharmonicity as $\epsilon_- - \epsilon_{\tilde{f}} = \Omega_R + \tilde{\alpha}$.
At this stage, we turn to numerics and construct a unitary basis transformation between the undriven and driven transmon eigenstates for a finite truncation of the transmon Hilbert space.
We then re-express the cross-Kerr interaction in the driven basis, giving us:
\be
\hat{H}/\hbar = \sum_i \epsilon_i \ket{i}\bra{i} + \hat{H}_c/\hbar -\chi\hat{c}^\dagger\hat{c}\sum_{jk}u_{jk}\ket{j}\bra{k}
\ee
We can further simplify this by going into the frame of the driven transmon $\hat{U} = e^{i\hat{H}_\textrm{Rabi}t/\hbar}$ which performs the transformations $\ket{j} \rightarrow e^{i\epsilon_jt/\hbar}\ket{j}$, resulting in:
\be
\hat{H}/\hbar = \hat{H}_c/\hbar -\chi\hat{c}^\dagger\hat{c}\underbrace{\sum_{jk}u_{jk}e^{i(\epsilon_j-\epsilon_k)t/\hbar}\ket{j}\bra{k}}_{\hat{q}^\dagger\hat{q}}
\ee
where $u_{jk} = u^*_{kj}$. We consider the terms associated with the lowest three levels explicitly:
\begin{align}
\begin{split}
\hat{q}^\dagger\hat{q} = \,\, & u_{++}\ket{+}\bra{+} + u_{--}\ket{-}\bra{-} + u_{\tilde{f}\tilde{f}}\tilde{\ket{f}}\tilde{\bra{f}} + u_{+-}e^{i\Omega_Rt}\ket{+}\bra{-} + u_{-+}e^{-i\Omega_Rt}\ket{-}\bra{+} \\
& + u_{-\tilde{f}}e^{i(\Omega_R + \tilde{\alpha})t}\ket{-}\tilde{\bra{f}} + u_{\tilde{f}-}e^{-i(\Omega_R + \tilde{\alpha})t}\tilde{\ket{f}}\bra{-} + u_{+\tilde{f}}e^{i(2\Omega_R + \tilde{\alpha})t}\ket{+}\tilde{\bra{f}} + u_{\tilde{f}+}e^{-i(2\Omega_R + \tilde{\alpha})t}\tilde{\ket{f}}\bra{+}
\end{split}
\end{align}
At this stage, we pause and turn to simplify $\hat{H}_c$.
First, we choose to parameterize the two drive frequencies $\omega_{1/2} = \omega_c - \Delta_c \mp \Omega_R$.
By going into the rotating frame at the average of the drive frequencies $\hat{U} = e^{i(\omega_c - \Delta_c)t\hat{c}^\dagger\hat{c}}$, we arrive at:
\be
\hat{H}/\hbar = \hat{H}_\textrm{int}/\hbar + \Delta_c\hat{c}^\dagger\hat{c} - \varepsilon_1(\hat{c}e^{-i\Omega_R t + i\varphi_1} + \hat{c}^\dagger e^{i\Omega_R t - i\varphi_1}) - \varepsilon_2(\hat{c}e^{i\Omega_R t + i\varphi_2} + \hat{c}^\dagger e^{-i\Omega_R t - i\varphi_2})
\ee
Finally, we assume that the drive strengths are equal $\varepsilon_2 = -\varepsilon_1 = \varepsilon$ and parameterize the drive phases as their sum and differential components $\varphi_\Sigma = (\varphi_1 + \varphi_2)/2$ and $\varphi_\delta = (\varphi_1 - \varphi_2)/2$.
This allows us to further simplify our Hamiltonian to:
\be
\hat{H}/\hbar = \hat{H}_\textrm{int}/\hbar + \Delta_c\hat{c}^\dagger\hat{c} - 2i\varepsilon\textrm{sin}(\Omega_R t - \varphi_\delta)(\hat{c}e^{i\varphi_\Sigma} - \hat{c}^\dagger e^{-i\varphi_\Sigma})
\ee
We can observe here that the sum phase of the two sidebands contributes simply as a static rotation of $\hat{c}$, therefore we can always align to this frame by experimentally adjusting this phase.
Hence, we will set $\varphi_\Sigma = 0$ here on out to simplify our expressions.
At this stage, we aim to eliminate this time-dependent drive term by performing a displacement transformation $\hat{U} = e^{\xi(t)\hat{c}^\dagger - \xi^*(t)\hat{c}}$.
This is achieved by choosing $\xi(t) = \frac{2\varepsilon}{\Omega_R}\textrm{cos}(\Omega_R t + \varphi_\delta) = \xi_0(e^{i(\Omega_R t + \varphi_\delta)} + e^{-i(\Omega_R t + \varphi_\delta)})$ where $\xi_0 = \frac{\varepsilon}{\Omega_R}$, which also transforms $\hat{c} \rightarrow \hat{c}$ + $\xi(t)$.
This gives:
\begin{align} \label{eq:1}
\begin{split}
\hat{H}/\hbar & = \Delta_c(\hat{c}^\dagger + \xi^*(t))(\hat{c} + \xi(t)) - \chi(\hat{c}^\dagger + \xi^*(t))(\hat{c} + \xi(t))\sum_{jk}u_{jk}e^{i(\epsilon_j-\epsilon_k)t/\hbar}\ket{j}\bra{k} \\
& = \Delta_c(\hat{c}^\dagger\hat{c} + \xi(t)(\hat{c} + \hat{c}^\dagger) + \xi^2_0) - \chi(\hat{c}^\dagger\hat{c} + \xi_0(e^{i(\Omega_R t + \varphi_\delta)} + e^{-i(\Omega_R t + \varphi_\delta)})(\hat{c} + \hat{c}^\dagger) + \xi^2_0)\sum_{jk}u_{jk}e^{i(\epsilon_j-\epsilon_k)t/\hbar}\ket{j}\bra{k}
\end{split}
\end{align}
By substituting the expansion for $\hat{q}^\dagger\hat{q}$, discarding terms that rotate at $\Omega_R$ and higher, and neglecting constant offsets, we are left with an effective static interaction Hamiltonian:
\be \label{eq:3}
\hat{H}/\hbar = \Delta_c\hat{c}^\dagger\hat{c} -\chi \xi_0 u_{+-} (e^{-i\varphi_\delta}\ket{+}\bra{-} + e^{i\varphi_\delta}\ket{-}\bra{+})(\hat{c} + \hat{c}^\dagger) - \chi\hat{c}^\dagger\hat{c}(u_{++}\ket{+}\bra{+} + u_{--}\ket{-}\bra{-} + u_{\tilde{f}\tilde{f}}\tilde{\ket{f}}\tilde{\bra{f}})
\ee
Importantly, this approximation requires larger Rabi frequencies as we drive harder to induce larger desired interaction strengths (see subsection A. for a more detailed analysis).
Finally, by neglecting the final term (see subsection B.), we arrive at the conditional displacement Hamiltonian between the transmon and one cavity mode:
\be \label{eq:2}
\hat{H}/\hbar = \Delta_c\hat{c}^\dagger\hat{c} - g(\textrm{cos}(\varphi_\delta)\hat{\sigma}_x + \textrm{sin}(\varphi_\delta)\hat{\sigma}_y)(\hat{c} + \hat{c}^\dagger)
\ee
where $g = \chi\xi_0 u_{+-} \approx \frac{\chi\xi_0}{2}$ and we have defined our Pauli operators such that $\hat{\sigma}_z = \ket{+}\bra{+} - \ket{-}\bra{-}$.
Here, we have formally identified the qubit that we will use within the larger driven transmon Hilbert space.
As we can see, the coupling axis of the qubit fully depends on the differential phase $\varphi_\delta$ of the cavity sidebands relative to the qubit phase (which we have defined as zero here), which can be easily adjusted experimentally without invoking additional Hamiltonian terms.
Note that throughout our derivations, we have assumed no accidental frequency collisions that bring unintended Hamiltonian terms into resonance.

This scheme extends relatively straightforwardly to multiple cavity modes dispersively coupled to the same transmon.
Incorporating a pair of sidebands on each additional cavity is sufficient to activate a conditional displacement involving the driven transmon, as long as the resonance condition is satisfied.
It is also worth noting that we have assumed that each sideband couples only to a single cavity mode - in practice, finite crosstalk complicates the calibration procedure of activating all sidebands together.
In our experiment, we enact the aforementioned Hamiltonian Eq. (\ref{eq:2}) for two cavity modes $\hat{c} \in \{ \hat{a}, \hat{b} \}$ coupled to orthogonal axes of the qubit.

\subsection{Optimizing the static cross-Kerr}
Here, we consider (Eq. \ref{eq:1}) and ask the questions: How large do we need the Rabi frequency to be in order to safely discard all the rotating terms? Does the answer to this question inform any design choices with regard to our static Hamiltonian?
To answer this question, we consider all of the different terms that rotate at $\Omega_R$, neglecting any phases:
\begin{align*}
\big( \Delta_c\xi_0(\hat{c} + \hat{c}^\dagger) -\chi u_{+-}\hat{c}^\dagger\hat{c}\hat{\sigma}_\mp - \chi \xi^2_0 u_{+-} \hat{\sigma}_\mp\big)e^{\pm i\Omega_R t}
\end{align*}
Note that there are also terms that rotate at $\tilde{\alpha}, 2\Omega_R, \Omega_R + \tilde{\alpha}, 2\Omega_R + \tilde{\alpha}$, and $3\Omega_R + \tilde{\alpha}$, but since the prefactors will all be of the same order, we consider the smallest rotating frequency for the most stringent condition.
Importantly, we also require that $\tilde{\alpha} > \Omega_R$, otherwise other terms involving $\tilde{\ket{f}}$ will be activated and we can no longer restrict ourselves to a qubit subspace.
This sets a limit on how large of a Rabi frequency can be used for a fixed anharmonicity $\alpha_q$.
The above terms reveal that our conditions for the RWA are:
\be 
\Omega_R \gg \{ \Delta_a |\xi_0 \langle(\hat{c} + \hat{c}^\dagger)\rangle|, \frac{\chi}{2}\langle\hat{c}^\dagger\hat{c}\rangle, \frac{\chi}{2}\xi^2_0 \}
\ee
which notably depends on the state of the cavity.
It is clear from this that as the conditional displacement interaction strength $g \approx \frac{\chi\xi_0}{2}$ increases, the approximation becomes less valid.
However, we can instead rewrite the condition for a fixed $g$:
\be 
\Omega_R \gg \{ \Delta_a |\xi_0 \langle(\hat{c} + \hat{c}^\dagger)\rangle|, \frac{g}{\xi_0}\langle\hat{c}^\dagger\hat{c}\rangle, g\xi_0 \}
\ee
which reveals that there is indeed an optimal value for $\xi_0$ given a fixed $g$. For considering photon numbers $\langle\hat{c}^\dagger\hat{c}\rangle \sim \mathcal{O}(1)$ and $g \approx \Delta_c$, we best satisfy all these conditions by choosing $\xi_0 \approx 1$.
This, in turn for a fixed $g$, suggests that we should roughly target a static cross-Kerr strength of $\chi \approx 2g$.
\subsection{Choice of static detuning of the Rabi drive}

The final term in Eq. (\ref{eq:3}) represents an effective cross-Kerr interaction between cavity photons and the driven transmon eigenstates.
For a true two-level system driven on resonance, which is a good approximation for transmons in the regime that the Rabi frequency is much weaker than the anharmonicity $\Omega_R \ll \alpha_q$, one finds that $u_{++} = u_{--} = 1/2$ which results in a static frequency shift of the cavity and hence a nulled cross-Kerr.
As the Rabi frequency $\Omega_R$ approaches the anharmonicity $\alpha_q$, however, $u_{++} \neq u_{--}$ for a drive that is on resonance owing to the hybridization of the driven eigenstates with higher energy levels of the transmon.
This results in a residual cross-Kerr which can be interpreted as a slight shift in the Rabi frequency due to the presence of photons in the cavity.
This is problematic as it both changes the resonance condition of the interaction and biases our measurement scheme as a function of the cavity photon distribution.

\begin{figure*}[t]
  \includegraphics[scale=0.4]{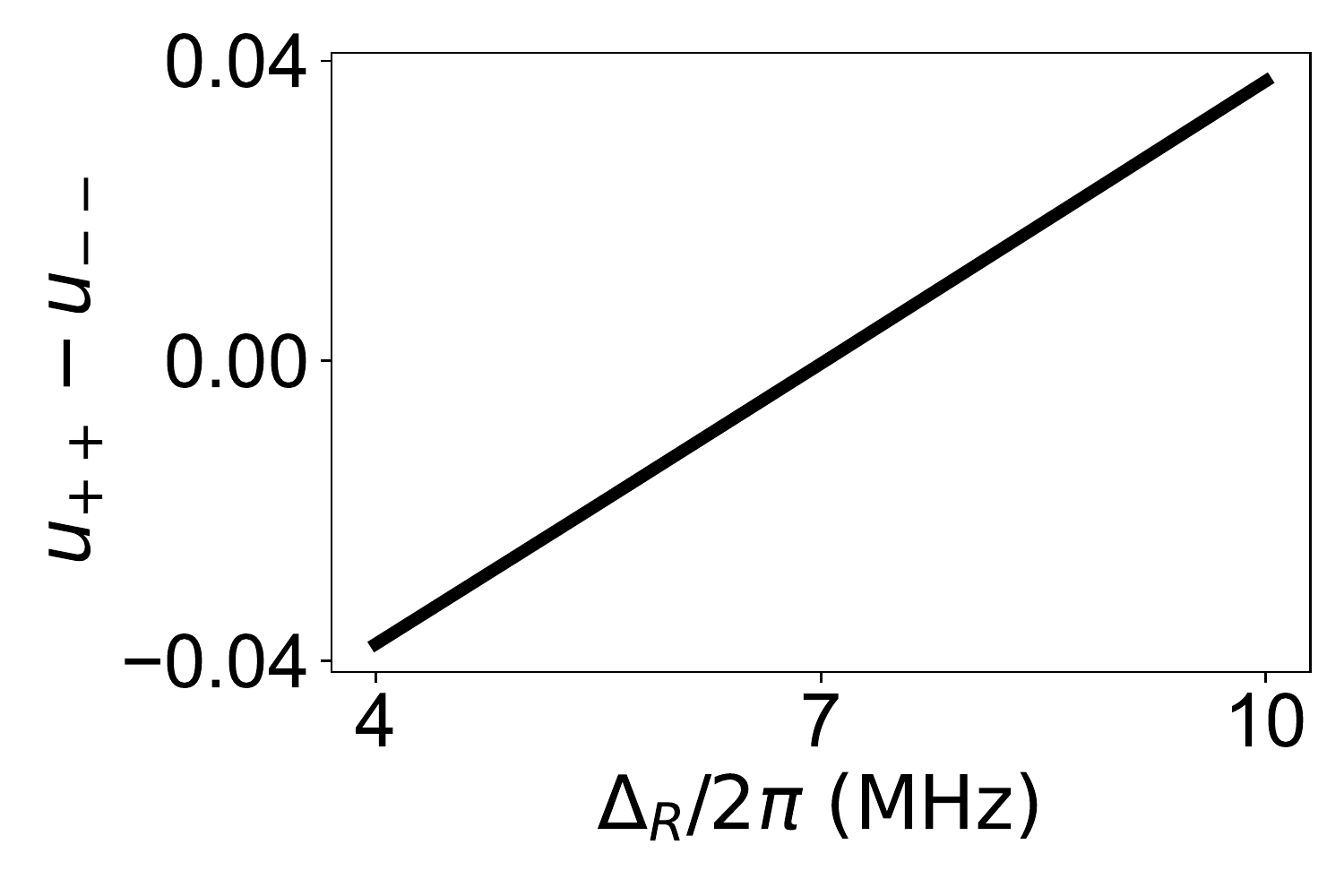}
  \caption{\textbf{Optimizing the static detuning.} By numerically diagonalizing $\hat{H}_q/\hbar$ for various values of $\Delta_R$, we can plot the dimensionless factor that contributes to the residual cross-Kerr for $\Omega_R/2\pi = 80$ MHz and $\alpha_q/2\pi = 244$ MHz. We find an optimal value of $\Delta_R/2\pi \approx$ 7 MHz.}
  \label{fig:fig1}
\end{figure*}

By adding an additional static detuning knob $\Delta_R$ on the Rabi drive, we can determine an optimal working configuration that nulls this effective cross-Kerr.
We show this optimization in FIG. \ref{fig:fig1}.
The presence of this static detuning thus dictates that we perform an adiabatic preparation of our driven qubit eigenstates.
This has the further benefit of eliminating leakage events associated with large Rabi frequencies and finite transmon anharmonicity, but has an additional challenge which we address in Section V. D.

\section{Measuring the driven qubit}

In order to measure the qubit along its original driven basis, i.e. $\hat{\sigma}_x$, we need to know the total phase accumulated on the excited state in the lab frame in order to perform the final $\pi/2$ rotation (in order to decode onto our measurement basis) around the correct axis.
As depicted in Fig. 2b of the main text, this phase should be equal to the Rabi frequency multiplied by the total free evolution duration plus an additional offset phase that is fixed for a given set of ramp times and drive amplitudes.
Since our decoding rotation is performed with a local oscillator that is locked to the $\ket{g} \leftrightarrow \ket{e}$ transition frequency of the undriven transmon, the proper phase to perform this rotation is $(\Omega_R - \Delta_R)\tau + \varphi_0$, which notably depends on time.

\begin{figure*}[t]
  \includegraphics[scale=0.9]{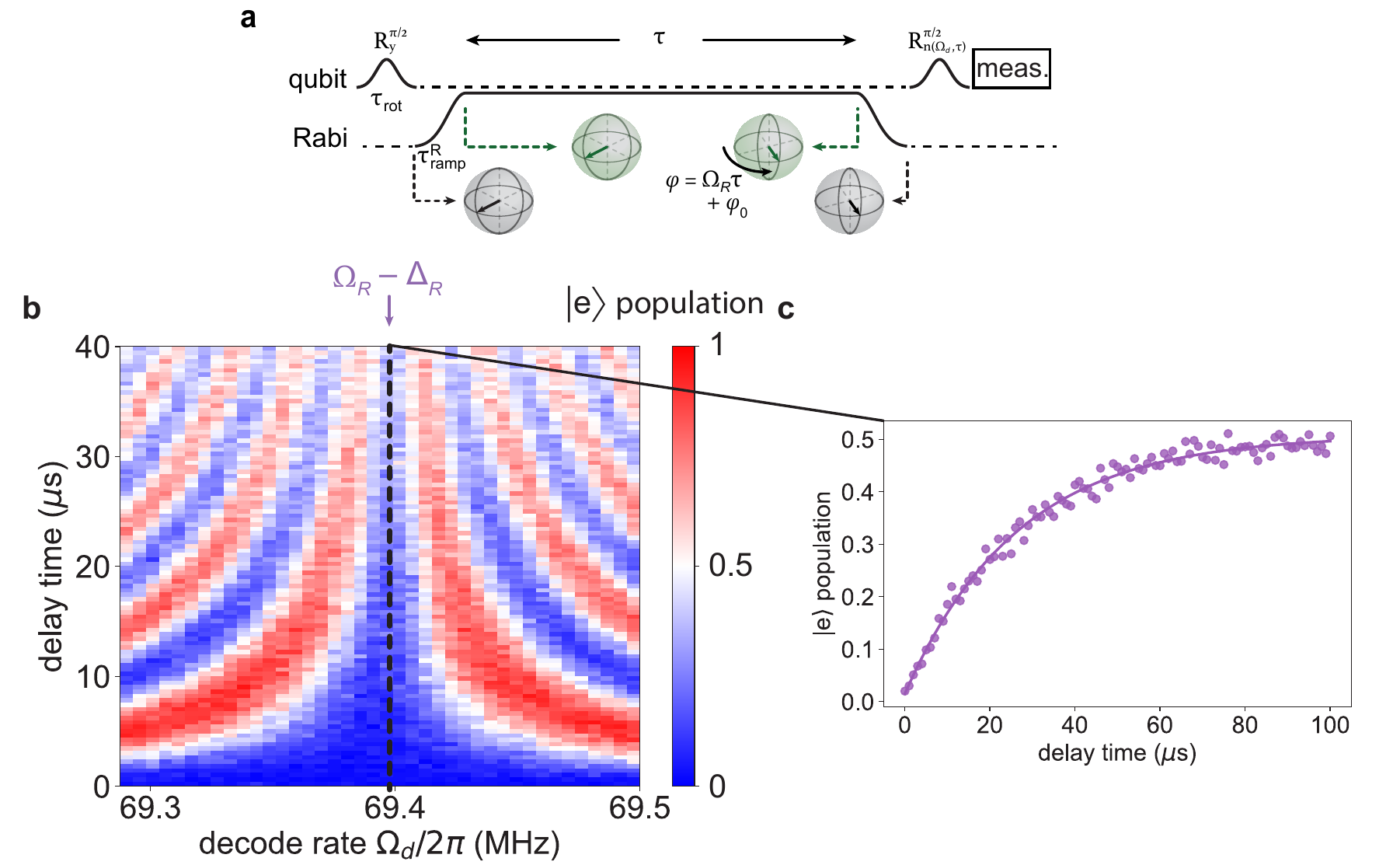}
  \caption{\textbf{State tomography and decoherence of transverse Bloch vectors. a.} A Ramsey-style pulse sequence is used to calibrate the decode rate. The final $\pi/2$ rotation has a phase that depends on both the delay time $\tau$ and the programmed decode rate $\Omega_d$. \textbf{b.} The 2D plot reveals an optimal decode rate which is inferred to be equal to the true Rabi frequency offset by the static detuning $\Omega_R - \Delta_R$. \textbf{c.} The decode rate can then be fixed to the optimal value to extract a driven transverse relaxation time $T_{2\rho} \approx$ 27 $\mu$s via fitting to an exponential decay function (solid line). }
  \label{fig:fig2}
\end{figure*}

FIG. \ref{fig:fig2} shows a calibration experiment where we scan the decode rate and the delay time in order to reveal the Rabi frequency (offset by the static detuning), which can then be fit to extract a driven transverse decoherence time $T_{2\rho}$ \cite{Gustavsson2012}.

\newpage
\section{Experimental Setup and Device Parameters}

\begin{figure*}[!]
  \includegraphics[scale=0.8]{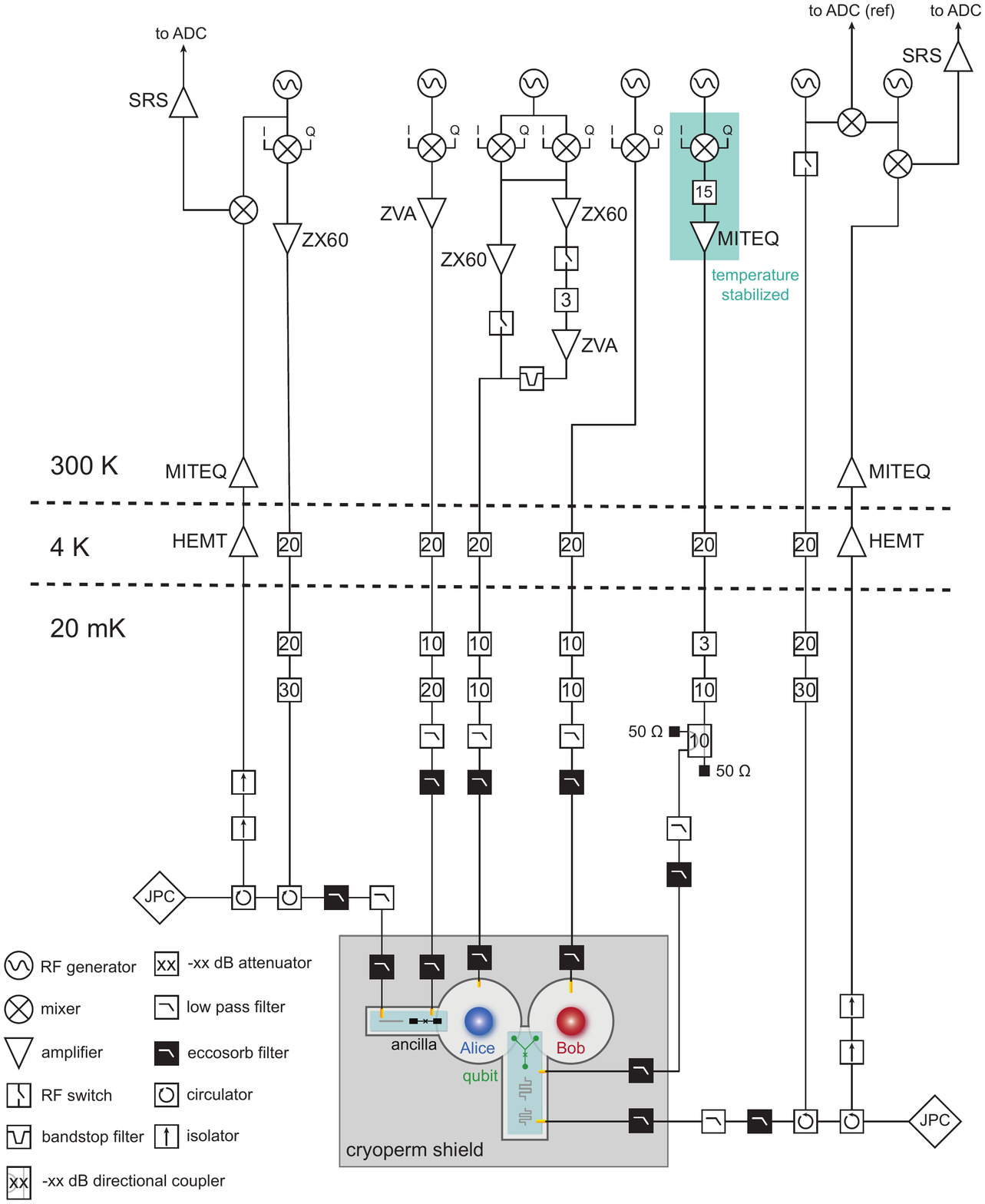}
  \caption{\textbf{Experimental wiring diagram.} }
  \label{fig:fig3}
\end{figure*}

A table of the static system parameters is given in TABLE I. 
A schematic of the wiring diagram for this experiment is depicted in FIG. \ref{fig:fig3}.
We comment below (subsection A.) on notable features not explicitly shown in the wiring diagram.

\subsection{Wiring Diagram}

An FPGA-based quantum controller synchronizes multiple modules that contain DACs and ADCs for generating the pulses (I \& Q control) and digitizing the readout signals, respectively.
RF switches are only open while pulses are played on a corresponding mode.
The control line for Alice is split between the resonant drive (left) and sideband pumps (right), which are never played simultaneously (see Fig. 2b of the main text), thus a closed loop where both RF switches are open together is never formed.
The bandstop filter on the pump line is centered on Alice's resonance frequency, suppressing pump-induced noise that may lead to heating of the cavity mode.
DC blocks are placed around each active component, as well as on each line at the boundary of the cryostat (i.e. the line separating 300 K and 4 K in the schematic).
All components in the 20 mK region are thermally anchored to the mixing chamber plate via OFHC copper braids.
We use Josephson Parametric Converters (JPCs) as quantum limited amplifiers - only coupling to the signal port is shown.

\renewcommand{\arraystretch}{1}
\begin{table*}[]
\setlength{\tabcolsep}{8pt}
\begin{tabular}{|c|c|c|}\hline
\multicolumn{1}{|c|}{\textbf{System quantity}} & \textbf{Parameter} & \textbf{Value}  \\ \hline
\multicolumn{1}{|c|}{Transmon frequency} & $\omega_q/2\pi$ & 4850 MHz \\ \hline
\multicolumn{1}{|c|}{Transmon anharmonicity} & $\alpha_q/2\pi$ & 244 MHz \\ \hline
\multicolumn{1}{|c|}{Transmon relaxation} & $T^q_1$ & 80 $\mu$s \\ \hline
\multicolumn{1}{|c|}{Transmon decoherence} & $T^q_2$ & 7 $\mu$s \\ \hline
\multicolumn{1}{|c|}{Alice frequency} & $\omega_a/2\pi$ & 5436 MHz \\ \hline
\multicolumn{1}{|c|}{Alice linewidth} & $\kappa_a/2\pi$ & 0.23 kHz \\ \hline
\multicolumn{1}{|c|}{Bob frequency} & $\omega_b/2\pi$ & 6506 MHz \\ \hline
\multicolumn{1}{|c|}{Bob linewidth} & $\kappa_b/2\pi$ & 320 kHz \\ \hline
\multicolumn{1}{|c|}{Alice - Transmon coupling} & $\chi_{aq}/2\pi$ & 295 kHz \\ \hline
\multicolumn{1}{|c|}{Bob - Transmon coupling} & $\chi_{bq}/2\pi$ & 210 kHz \\ \hline
\multicolumn{1}{|c|}{Ancilla frequency} & $\omega_{qa}/2\pi$ & 4509 MHz \\ \hline
\multicolumn{1}{|c|}{Ancilla relaxation} & $T^{qa}_1$ & 60 $\mu$s \\ \hline
\multicolumn{1}{|c|}{Ancilla decoherence} & $T^{qa}_2$ & 10 $\mu$s\\ \hline
\multicolumn{1}{|c|}{Alice - Ancilla coupling} & $\chi_{a,qa}/2\pi$ & 845 kHz\\ \hline

\end{tabular}
\caption{$\textbf{List of system parameters.}$ }
\end{table*}

\subsection{Temperature stabilization}

\begin{figure*}[t]
  \includegraphics[scale=0.8]{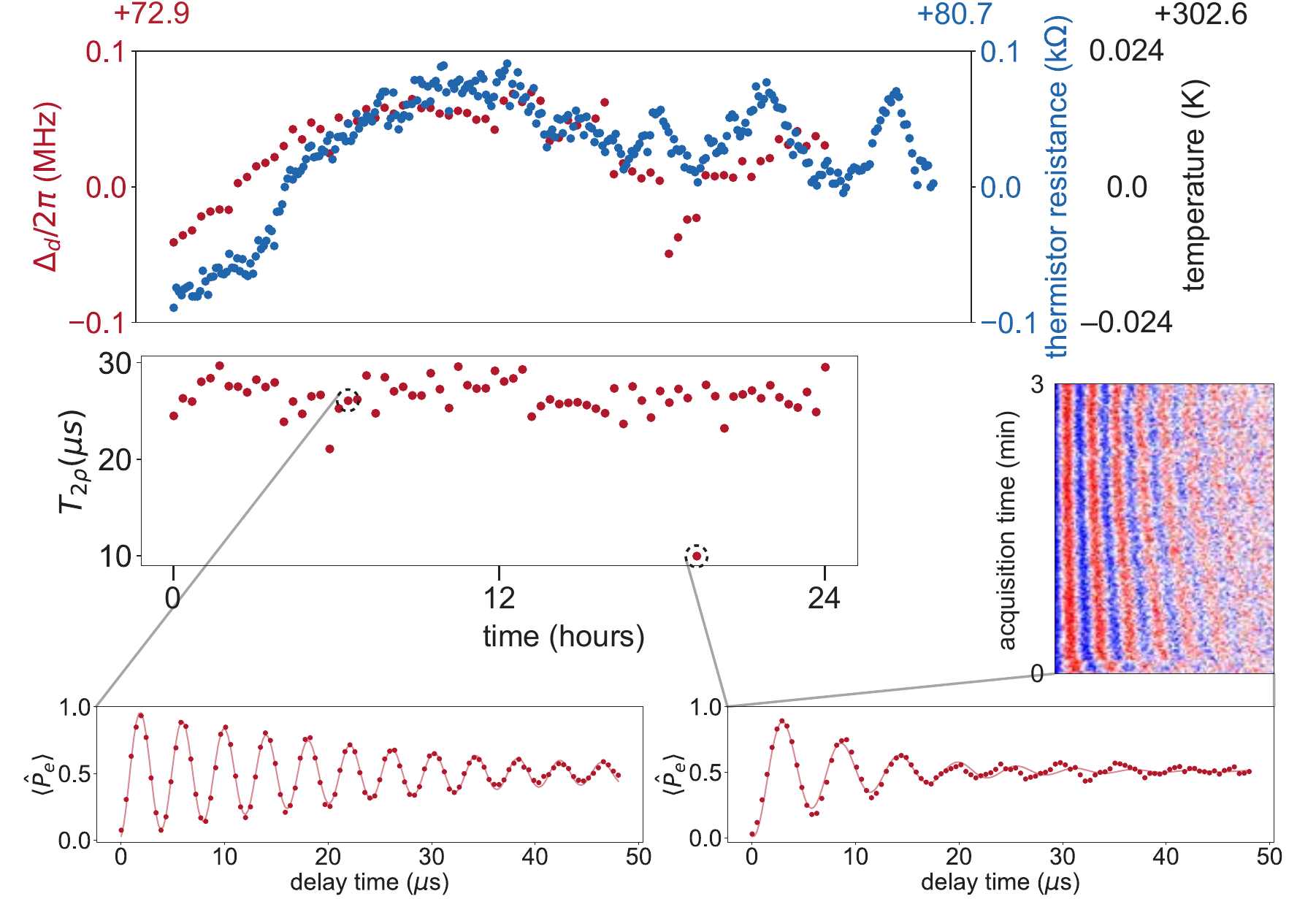}
  \caption{\textbf{Tracking system stability over time.} (Top panel) Simultaneous measurements of the Rabi frequency and the ambient temperature where the active microwave components are held reveals correlations between temperature drifts and amplitude drifts. (Middle panel) Extracting driven coherences $T_{2\rho}$ suggests a stable amplitude noise spectrum within an acquisition time $\tau_\textrm{acq}$ = 3 minutes (bottom left panel: a typical time-domain Ramsey trace), with a single instance where the amplitude drift was large (bottom right panel), as confirmed via looking at the raw data binned 10 shots at a time over $\tau_\textrm{acq}$.}
  \label{fig:fig4}
\end{figure*}

As the derivations of Section II. suggest, the resonance condition for enacting our reaction Hamiltonian (Eq. (3) of the main text) relies on matching the sideband detunings to the Rabi frequency.
The Rabi frequency depends linearly on the amplitude of the Rabi drive (roughly speaking, using the two-level approximation that $\Omega_R = \sqrt{\epsilon^2_R + \Delta^2_R}$ and we operate in a regime where $\epsilon_R \gg \Delta_R$, resulting in $\Omega_R \approx \epsilon_R(1 + \frac{\Delta^2_R}{2\epsilon^2_R})$), and thus is susceptible to amplitude fluctuations such as those caused by variations in the gain at any stage of our microwave control chain.
A dominant source of these variations is due to ambient temperature fluctuations in the lab.
To this end, we suppress these fluctuations by anchoring the components (particularly a Marki LXP IQ-mixer and MITEQ low noise amplifier, see turquoise box in FIG. \ref{fig:fig3}) to a Thorlabs optical breadboard and placing the breadboard in a cardboard box.
We then actively stabilize the temperature of the air in the box via an op-amp based PID feedback controller that heats an Ohmite ceramic resistor ($R$ = 2.5 $\Omega$) based on a differential measurement of the temperature using a 100 k$\Omega$ thermistor referenced to a set point.
FIG. \ref{fig:fig4} shows the typical performance of our stabilization and correlates the temperature variations with the amplitude variations as measured via the Rabi frequency.
Over the course of 24 hours, we achieve an absolute temperature stability within 50 mK and a relative amplitude stability of 100 kHz / 72.9 MHz $\approx 10^{-3}$, suggesting that we have a relative amplitude sensitivity of 1\% per 500 mK.
We note that the timscale for a typical calibration and measurement of a dataset presented in this paper is roughly a few hours, meaning we can operate in a window where the relative amplitude stability can be much better than $10^{-3}$.

\section{Calibration Procedures}

In this section, we provide an overview on the calibration procedures in our experiment.
We calibrate the interaction between each individual cavity sideband and our driven qubit before proceeding to combine sidebands to enact conditional displacements.
As more drives are added, care is taken to re-optimize the resonance conditions given that the presence of each additional drive will result in a Stark shift of each mode, particularly the transmon whose Rabi frequency we are very sensitive to.

\subsection{Individual sideband interactions}

The sideband interactions have the form:
\begin{align}
\hat{H}_\textrm{red}/\hbar & = g\hat{c}\hat{\sigma}^-_x + g^*\hat{c}^\dagger\hat{\sigma}^+_x  \\
\hat{H}_\textrm{blue}/\hbar & = g\hat{c}\hat{\sigma}^+_x + g^*\hat{c}^\dagger\hat{\sigma}^-_x
\end{align}
for a general cavity annihilation operator $\hat{c}$ and $\hat{\sigma}^\pm_x = \ket{\mp}\bra{\pm}$ are the raising and lowering operators of the two driven qubit eigenstates that adiabatically connect to the ground and first excited state of the transmon. 
For our system, we have $\hat{c} \in \{\hat{a}, \hat{b}\}$.
The qualitative behavior of each individual sideband interacting with the driven qubit will be different given that we are operating in the regime where $g > \kappa_a$ and $g \lesssim \kappa_b$.
The former will result in either creating and annihilating two excitations simultaneously $\ket{+,0} \leftrightarrow \ket{-,1}$ (red sideband) or a coherent exchange between an excitation in the driven qubit and a photon in the cavity $\ket{-,0} \leftrightarrow \ket{+,1}$ (blue sideband) \cite{Lu2017}.
The latter will stabilize the qubit in either the excited state $\ket{-}$ (red sideband) or the ground state $\ket{+}$ (blue sideband) \cite{Murch2012}.

In order to calibrate the interaction strengths, we will operate in the restricted two-dimensional subspace of the joint Hilbert space of the cavity and qubit as described above. 
This allows us to simplify our analysis and replace the qubit raising and lowering operators $\hat{\sigma}^\pm_x$ with general bosonic creation and annihilation operators $\hat{d}^{(\dagger)}$.
We then follow Refs. \cite{Pfaff2017, Grimm2020} and capture the full range of dynamics by solving the equations of motion for $\hat{d}$ under $\hat{H}_\textrm{blue}$ and incorporating a cavity damping rate $\kappa$.
We also include a static detuning term $\delta\hat{c}^\dagger\hat{c}$ to capture the effect of sweeping the pump frequency that enables the interaction.
The resulting field has the form:
\be \label{eq:4}
\hat{d}(t) = \frac{\hat{d}(0)}{\Omega}e^{-\frac{\kappa_\textrm{eff}t}{4}}\bigg(\Omega\textrm{cosh}\big(\frac{\Omega t}{4}\big) + \kappa_\textrm{eff}\textrm{sinh}\big(\frac{\Omega t}{4}\big)\bigg)
\ee
where $\Omega = \sqrt{\kappa^2_\textrm{eff} - (4g)^2}$ and $\kappa_\textrm{eff} = \kappa + 2i\delta$.

\begin{figure*}[t]
  \includegraphics[scale=0.8]{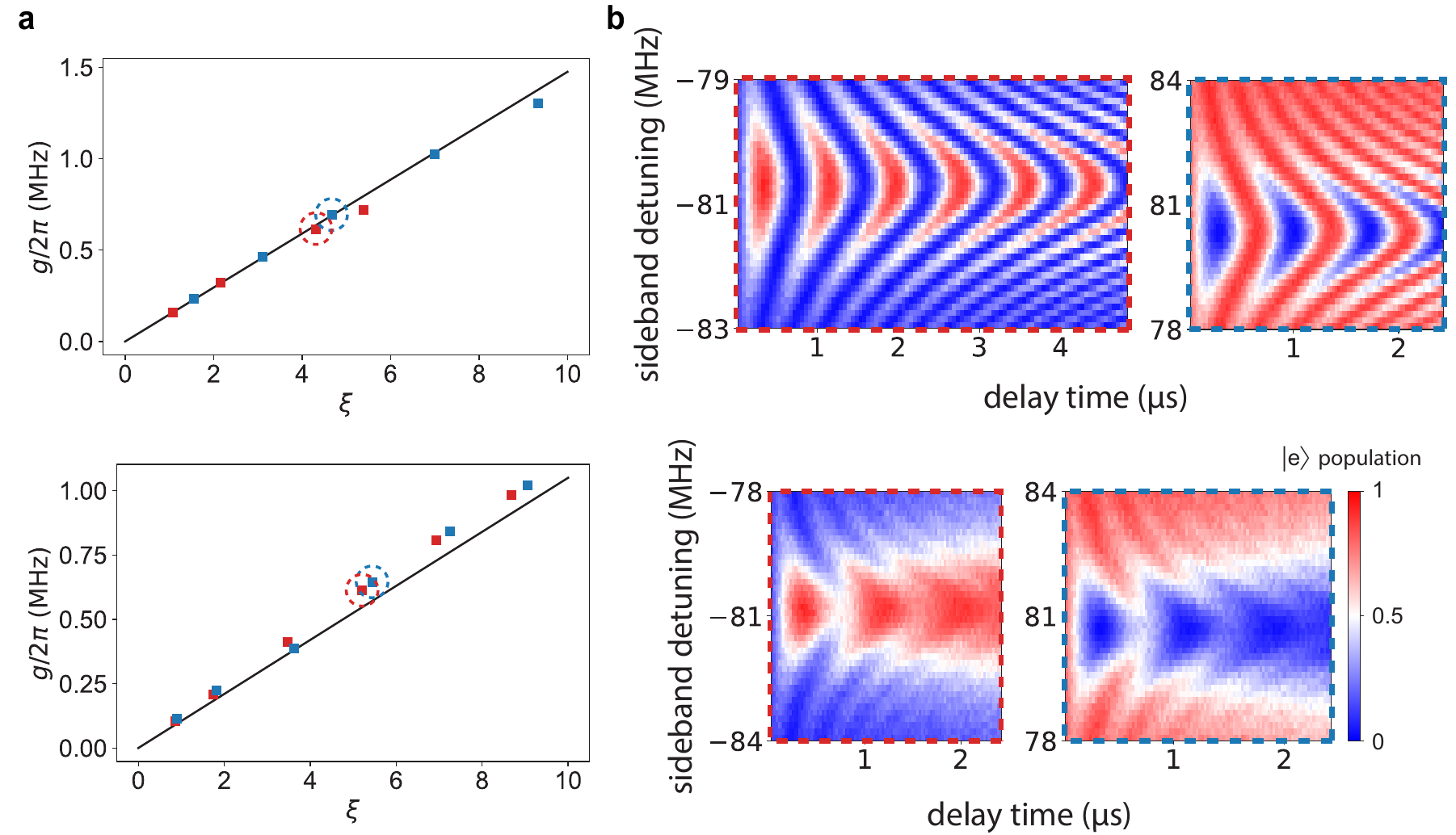}
  \caption{\textbf{Calibrating individual exchange interactions a.} Interaction strengths $g$ for individual sideband interactions between the driven qubit and Alice (top panel) or Bob (bottom panel). $g$ is extracted via a fit using Eq. (\ref{eq:4}) for the red sidebands (red squares) and blue sidebands (blue squares). Error bars are smaller than the markers. $\xi$ is calibrated via independent measurements of $\chi_{a/b}$ and the qubit's Stark shift as a function of pump amplitude. Solid lines represent the expected interaction strength in the ideal case $g = \frac{\chi_{a/b} \xi}{2}$. Deviations of the data from a linear relation for Alice coincides with regions where the Stark shift becomes nonlinear in pump power. Slight disagreement between the data and theory expectation for Bob may stem from an inaccurate estimate of $\chi_b$ and from a sensitivity of the extracted fit values to a true value of $\kappa_b$. \textbf{b.} Raw calibration data of the qubit population for a fixed pump amplitude (dashed circles in a.) for Alice (top) and Bob (bottom). The red sideband (left) either coherently creates and annihilates correlated photons in the driven qubit and the resonator ($g > \kappa_a$, Alice) or stabilizes the driven qubit in the first excited state ($g \lesssim \kappa_b$, Bob). The opposite is true for the blue sideband (right), which either coherently exchanges a photon between the driven qubit and resonator (Alice) or stabilizes the driven qubit in the ground state (Bob). Simultaneous measurement of the photon population in Alice via its ancilla qubit reveals the aforementioned correlations. The sideband detunings are referenced to the cavity frequencies, showing that we are operating at Rabi frequencies $\Omega_R/2\pi \approx 80$ MHz.}
  \label{fig:fig5}
\end{figure*}

For each interaction, we prepare our system in either $\ket{-,0}$ (blue sidebands) or $\ket{+,0}$ (red sidebands) and scan the frequency of the cavity sideband and the delay time for a given pump amplitude.
By measuring the qubit population, we extract $\langle\hat{d}^\dagger(t)\hat{d}(t)\rangle$ and can fit the resulting data using Eq. (\ref{eq:4}).
For the exchange interaction under the blue sideband, we have $\langle\hat{d}^\dagger(0)\hat{d}(0)\rangle_\textrm{blue} = 1$ for the qubit initially in its excited state, giving us an expression for $\langle\hat{d}^\dagger(t)\hat{d}(t)\rangle_\textrm{blue}$.
For the red sideband, the features are qualitatively identical, with the exception that the qubit starts out in the ground state, giving us $\langle\hat{d}^\dagger(t)\hat{d}(t)\rangle_\textrm{red} = 1 - \langle\hat{d}^\dagger(t)\hat{d}(t)\rangle_\textrm{blue}.$
Notably, this assumes that the effective interaction strength $g$ is independent of the pump detuning $\delta$, which is strictly not true but should be a very good approximation in our regime given that the scale of the chevron features, set by $g$, is much smaller than the absolute detuning from the cavity resonance $\sim\Omega_R$.
We allow for an overall amplitude, global offset, and time offset in our fit, leaving $g$ and $\kappa$ to be the only free parameters.
In the case of Bob, we first perform this fit for a range of interaction strengths $g \approx \kappa_b$ and extract a decay rate $\kappa_b/2\pi \approx 320$ kHz.
For the remainder of the calibrations where $g < \kappa_b$, we fix this quantity and let the interaction strength $g$ be the only free parameter to be fitted.
The full calibration for different pump amplitudes is shown in FIG. \ref{fig:fig5}.

\subsection{Calibrating conditional displacements}

The combination of simultaneous red and blue cavity sidebands enacts a conditional displacement interaction.
This requires that two conditions are fulfilled: 1) the interaction strengths of each individual sideband interaction be equal and 2) the frequency difference between the two sidebands equals twice the Rabi frequency.
If these two conditions are met, then we can model the interaction using Eq. (\ref{eq:1}).

In practice, the presence of each sideband will Stark shift both the transmon and cavity modes.
Thus, to capture the dominant effect of all of these Stark shifts (which influences the resonance condition), we perform individual sideband calibrations with the opposite sideband on but detuned by an amount larger than the interaction strength we are using (i.e. by an additional 2 MHz in our experiments).
We scan the pump amplitudes and match the individual sideband strengths before bringing both sidebands into resonance.
This relies on the assumption that over a variation of $\sim$2 MHz, the relative change in the cavity Stark shift, which influences the dimensionless pump strength that determines the interaction strength, is negligible.
Finally, we fine tune the difference frequency of the two sidebands while keeping the average value fixed (which fixes $\Delta_c$ in Eq. (\ref{eq:1})) in order to account for any change in the Rabi frequency which we are very sensitive to.
This sensitivity is revealed by measuring the transverse relaxation time $T_{2\rho}$, and choosing a calibration point where this value is maximized, suggesting that the resonance conditions are fulfilled as best as possible.
This calibration procedure gives us the data shown in Fig. 2 of the main text.

For Alice, the dynamics of an initial vacuum state $\ket{\alpha_0 = 0}_a$ evolving under a conditional displacement interaction will be a circular trajectory in phase space around the location of the ground state $\alpha_g = g_x/\Delta_a$.
By measuring the population in $\ket{n = 0}_a$, we are effectively measuring the overlap of a coherent state with itself as it oscillates in time.
This justifies the use of a simple model, where the state autocorrelation function is $\langle\beta e^{-i\Delta t}|\beta\rangle = e^{|\beta|^2(e^{-i\Delta_a t} -1)}$.
The corresponding probability is:
\be
P_0 = |\langle\beta e^{-i\Delta t}|\beta\rangle|^2 = e^{2|\beta|^2(\textrm{cos}[\Delta_a t] - 1)}
\ee
To make the connection with our model, we choose $\beta = \alpha_g = g_x/\Delta_a$.

\subsection{Calibrating cavity displacements along the interaction axis}

As described in Section II., the cavity phase of the conditional displacement (i.e. the phase which defines the position operator $\hat{x} \propto (\hat{c}e^{i\varphi_\Sigma} + \hat{c}^\dagger e^{-i\varphi_\Sigma})$) is determined by the sum phase of the red and blue sidebands.
Given that we are turning on the conditional displacement interaction suddenly ($\tau^\textrm{sb}_\textrm{ramp} \ll 1/g$), the phase of our initial displacement operation $\hat{\mathcal{D}}(\alpha_0)$ on Alice will determine the location in the driven phase space where the wave-packet begins.
Displacements whose phase is aligned to the conditional displacement cavity phase will prepare wave-packets along the position axis, whereas care needs to be taken to prepare wave-packets with various momentums that are located at one of the two ground state positions.
Since we are interested in modeling scenarios where a wave-packet arrives on a potential energy surface via optical excitation in a Franck-Condon region, we prepare coherent states with no initial momentum along the reaction coordinate.

In order to calibrate the displacement phase, we begin with a calibrated conditional displacement where we have extracted $g_x$ and $\Delta_x$, which gives us a value for the ground state amplitude $\alpha_g = g_x/\Delta_x$ (Fig. 2c of the main text).
Note that this does not rely on any displacement phase since we are starting off in a vacuum state.
Next, we scan the phase of an initial displacement of $2\alpha_g$ and turn on the conditional displacement interaction for various delay times.
The optimal phase will be the one where we recover revivals that are half a period out of phase from those in Fig. 2c of the main text.
This can be interpreted as follows. 
A vacuum state in the lab frame $\ket{0}_\textrm{lab}$ looks like a displaced state $\ket{-\alpha_g}_\textrm{disp}$ with respect to the displaced ground state $\ket{\alpha_g}_\textrm{lab} = \ket{0}_\textrm{disp}$, and thus will oscillate around the ground state, reaching $\ket{\alpha_g}_\textrm{disp}$ after half a period.
By determining the phase that enables us to prepare $\ket{\alpha_g}_\textrm{disp}$ (which is $\ket{2\alpha_g}_\textrm{lab}$ in the lab frame and will return to the vacuum state $\ket{-\alpha_g}_\textrm{disp}$ after half a period) we can prepare any state along the position axis, including the displaced ground state.

\begin{figure*}[!]
  \includegraphics[scale=1]{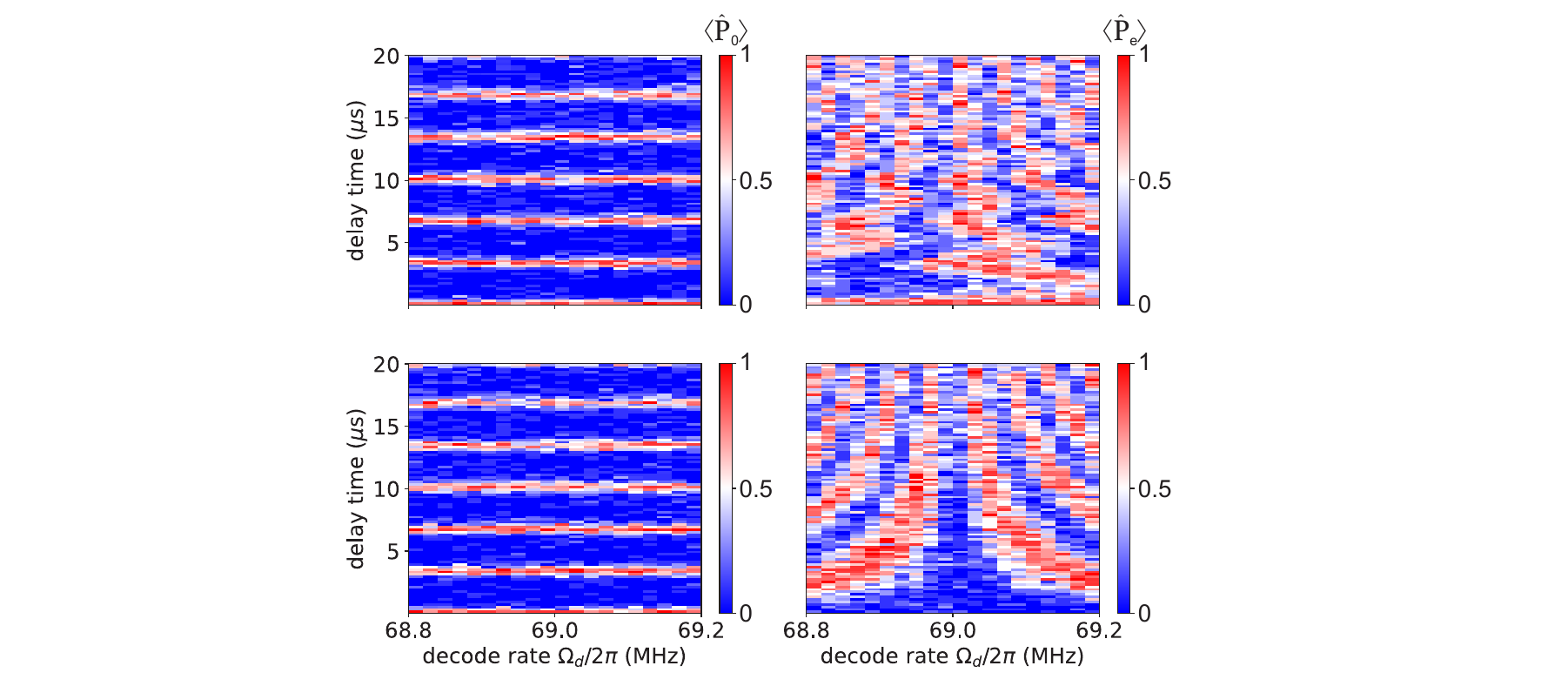}
  \caption{\textbf{Echoing away residual entanglement.} Simultaneous measurement of Alice's vacuum projector via a transmon ancilla (left) and the driven qubit (right) while a conditional displacement interaction is active. The pulse sequence used for this calibration is shown in Fig. 2b of the main text without (top) and with (bottom) the $\pi$ rotation and echo delay $\tau_\textrm{ED}$ and only sidebands on Alice. There is no initial displacement amplitude ($\alpha_0 = 0$). Time dynamics of Alice's vacuum projector reveals coherent revivals as expected (as in Fig. 2c of the main text), independent of the decode rate on the driven qubit. Scanning the decode rate without the echo sequence results in distortions to the driven Ramsey data that are correlated with the cavity photon distribution. Implementing the echo sequence eliminates this effect, suggesting that the systems remain unentangled at all delay times. Here, we use an optimized value of the delay time $\tau_\textrm{ED} = 144$ ns.}
  \label{fig:fig6}
\end{figure*}

\subsection{Echoing the residual entanglement during ramping of the Rabi drive}

The adiabatic preparation of our driven qubit eigenstates as motivated in Section II. B. has two benefits: for a fixed transmon anharmonicity, we can 1) use larger Rabi frequencies while cancelling the residual cross-Kerr and 2) avoid leakage events to higher transmon levels (up to natural heating rates of the dressed eigenstates) associated with a resonant drive. The primary consequence of this approach is the undesired interaction between the qubit and cavity photons during the ramp time of the Rabi drive. 

Qualitatively, this undesired interaction stems from ramping between the static interaction Hamiltonian $\hat{H}_\textrm{int}/\hbar = -\chi\hat{c}^\dagger\hat{c}\hat{q}^\dagger\hat{q}$ and the driven Hamiltonian where the cross-Kerr interaction is nulled.
Thus, if there are photons in the cavity during either the ramp on or off of the Rabi drive, they will entangle with the superposition states of the qubit that we are manipulating.
We can avoid this effect during the ramp on of the Rabi drive by performing our displacement operation after the Rabi drive is fully ramped on, i.e. during the time when the cross-Kerr interaction is nulled (hence the relative position of the displacement operation in Fig. 2b of the main text).
For addressing the entanglement during the ramp off of the Rabi drive, we implement a simple and short echo sequence of the qubit to un-do the interaction.
This works because the entanglement is fully determined by $\chi$ and the ramp time $\tau^R_\textrm{ramp}$ and not the cavity photon distribution.
This is important as we do not want a scheme which depends on the cavity state that we are manipulating.
FIG. \ref{fig:fig6} shows how implementing this protocol eliminates spurious features that arise from this entanglement when performing a decode calibration experiment (FIG. 2b) when a conditional displacement interaction with Alice is active.
In practice, this calibration is only done with respect to photons in Alice.
Given that we operate Bob in a regime where $g_y < \kappa_b$, the photon distribution in Bob remains relatively small and thus any residual entanglement effects are negligible.

\section{Extended data \& post-selecting on leakage events}

\begin{figure*}[!]
  \includegraphics[scale=1]{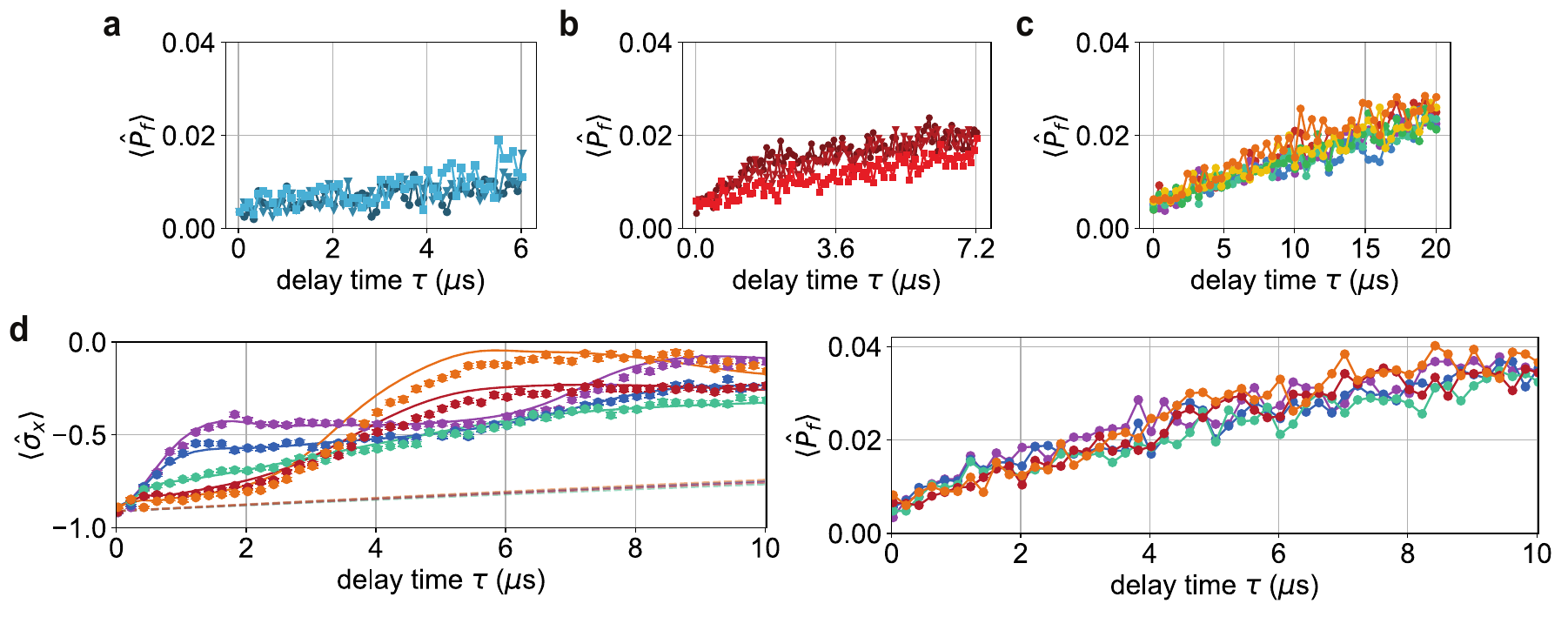}
  \caption{\textbf{Extended data and leakage statistics. a,b,c.} Probability of leaking out of the qubit subspace to $\ket{f}$ for the data presented in Figs. \{2b, 2c, 3b\} of main text, respectively. \textbf{d.} The same dataset as appears in Fig. 4b of the main text (left), but including two additional initial states with $\ket{\alpha_0}_a = \ket{\alpha_g/2}_a$ (blue) and $\ket{3\alpha_g/2}_a$ (red), with the associated leakage probabilities (right). All of the leakage probabilities are equivalent to the percentage of data that is post-selected away for the rest of the data in the paper, which is always less than 5\%.}
  \label{fig:fig7}
\end{figure*}

In this section we present additional data (FIG. \ref{fig:fig7}) that supports what is shown in the main text, specifically focusing on leakage statistics.

\renewcommand{\arraystretch}{1}
\begin{table*}[]
\setlength{\tabcolsep}{6pt}
\begin{tabular}{|c|c|c|c|c|c|}\hline
\multicolumn{1}{|c|}{\textbf{Dataset}} & \textbf{Drives} (+ Rabi) & \textbf{Figure} & $\Omega_R[\Delta_c]/2\pi$ (MHz) & $T_{2\rho}$ [$\alpha_0$] ($\mu$s) & \textbf{Leakage} \\ \hline
\multirow{3}{*}{Coherent state revivals} & \multirow{3}{*}{Alice sidebands} & \multirow{3}{*}{2c} & 82.532 [0.457] & \multirow{3}{*}{not measured} & \multirow{3}{*}{SM Fig. \ref{fig:fig7}a} \\
& & & 82.532 [0.355] & & \\
& & & 82.492 [0.246] & & \\ \hline
\multirow{3}{*}{Measurement-induced dephasing} & \multirow{3}{*}{Bob sidebands} & \multirow{3}{*}{2d} & 79.915 [0] & 36.6 & \multirow{3}{*}{SM Fig. \ref{fig:fig7}b} \\
& & & 79.9175 [0.4] & 40.5 & \\
& & & 79.9225 [0.8] & 36.3 & \\ \hline
\multicolumn{1}{|c|}{Coherent revivals} & \multirow{2}{*}{Alice \& Bob sidebands} & \multirow{2}{*}{3b} & \multirow{2}{*}{80.045} & \multirow{2}{*}{see Fig. 3b} & \multirow{2}{*}{SM Fig. \ref{fig:fig7}c} \\ 
+ state preparation & & & & & \\ \hline
\multirow{5}{*}{Conical intersection dynamics} & \multirow{5}{*}{Alice \& Bob sidebands} & \multirow{5}{*}{4b} & \multirow{5}{*}{81.013} & 51.7 [0]& \multirow{5}{*}{SM Fig. \ref{fig:fig7}d} \\
& & & & 51.1 [$\alpha_g/2$] & \\
& & & & 57.8 [$\alpha_g$] & \\
& & & & 53.9 [$3\alpha_g/2$] & \\
& & & & 48.8 [$2\alpha_g$] & \\ \hline

\end{tabular}
\caption{$\textbf{Extended data information.}$ Quantities in brackets correspond to identifiers within a dataset. Different values of $\Delta_c$, $c \in \{a, b\}$ require a fine-tuning calibration on the drive configuration to match Rabi frequency shifts, but preparing different initial states does not.}
\end{table*}

Our measurement of the transmon is able to distinguish between $\{ \ket{g}, \ket{e}, \ket{f} \}$ on a single-shot basis, which gives us information on leakage events outside the qubit manifold that we use for our experiments.
In the absence of decoherence, our adiabatic preparation scheme should ideally eliminate any leakage to the second excited state $\ket{f}$ and higher (assuming there are no accidental resonances induced by the drives).
In practice, any relaxation or heating between undriven transmon levels will lead to transitions between the driven eigenstates that have finite support across multiple undriven basis states.
In all of the data presented, we post-select away outcomes where the transmon is measured to be in $\ket{f}$.
We compile the post-selection statistics in FIG. \ref{fig:fig7}, and note that overall the leakage probabilities are small.

\section{Dissipation analysis and Zeno dynamics}

In this section we describe the oscillator-induced dissipation of the qubit resulting from the evolution under the master equation,
\begin{align}
    \dot{\hrho}=-i[\hat{H}_\textrm{Zeno},\hrho]+\kb \left(\htb\hrho\htb^\dag-\frac{1}{2}\htb^\dag\htb\hrho-\frac{1}{2}\hrho\htb^\dag\htb\right)\label{met}
\end{align}
with
\begin{align}
    \hat{H}_\textrm{Zeno}=E(x)\sx+\Delta_y\htb^\dag\htb+g\sz(\htb^\dag+\htb)\label{ht}
\end{align}
from Eq. (5) of the main text.
The dynamics of the system are effectively the same as Zeno dynamics of a driven qubit generated by the competition between $E(x)\sx$, which tries to lock the state of the qubit to an eigenstate of $\sx$, and $g\sz(\htb^\dag+\htb)$, which in combination with photon loss in the oscillator tries to project the state of the qubit onto an eigenstate of $\sz$. 
Note that here we choose the dissipation to be along $\sz$ as opposed to $\hat{\sigma}_y$.
The resulting dissipative dynamics is crucially dependent on the parameters $E(x),\Delta_y, g,\kb$.
While an exact analytic expression for the time-dependence of the qubit density matrix cannot be obtained in general, here we consider a few extreme parameter regimes which can be easily analyzed and provide a window to the vast range of qubit dynamics possible.

This simplest case is when $E(x)=0$. 
In this case, it is possible to use Fokker-Planck equations to calculate an effective dephasing rate. 
However, when $E(x)\neq 0$, this approach will fail to give an analytical expression for the dephasing rate. 
In this case, we will invoke additional constraints which will allow us to apply the Born-Markov approximation and derive an effective master equation for the qubit.

\subsection{$E(x)\neq 0$}
To get a simple intuitive understanding of the oscillator-induced dephasing when $E(x)\neq 0$, we consider two cases: (\textbf{a}) $\Delta_y=0$, $\kb \gg g$ and (\textbf{b}) $|2\Delta_y-E(x)|\ll |2\Delta_y+E(x)| $, $|2\Delta_y+E(x)|\gg\kb \gg |2\Delta_y-E(x)|,g$. 
\\
\\
Case (\textbf{a}): In this case, the oscillator mode can be eliminated using the standard Born-Markov approximation and a simple effective master equation for the qubit is obtained, 
\begin{align}
    \hrho_q=-i[\hat{H}_q,\hrho_q]+\kappa_q \left(\sz\hrho_q\sz-\hrho_q\right)\label{me1}
\end{align}
with $\hat{H}_q=E(x)\sx$, $\kappa_q=g^2\kb/(4E(x)^2+\kb^2/4)$. 
In this case, the state of the qubit will be an equal mixture of $\ket{+}$ and $\ket{-}$ or $\ket{g}$ and $\ket{e}$.
\\
\\
Case (\textbf{b}): In this case, the standard Born-Markov approximation yields the following master equation, 
\begin{align}
    \hrho_q&=-i[\hat{H}_q,\hrho_q]\nonumber\\
    &+\kappa_q \left(\ket{-}\bra{+}\hrho_q\ket{+}\bra{-}-\frac{1}{2}\ket{+}\bra{+}\hrho_q-\frac{1}{2}\hrho_q\ket{+}\bra{+}\right)\label{me2}
\end{align}
with $\hat{H}_q=E(x)\sx$, $\kappa_q=g^2\kb/((2E(x)-\Delta_y)^2+\kb^2/4)$. 
Unlike in case (\textbf{a}), here, the steady state of the qubit will be the pure state $\ket{+}$.
The difference between the master equations Eq.~\eqref{me1} and Eq.~\eqref{me2} can be understood by observing that when $|2\Delta_y+E(x)|\gg\kb \gg |2\Delta_y-E(x)|$, then the transition between $\ket{+}\otimes |0\rangle$ and $\ket{-}\otimes |1\rangle$ becomes more likely than that between $\ket{+}\otimes |1\rangle$ to $\ket{-}\otimes |0\rangle$ because the former, happening at frequency $2\Delta_y-E(x)$ lies within the bandwidth $\kb$. 
The first term in the tensor product refers to the state of the qubit and the second term refers to the vacuum $\ket{0}$ and single-photon Fock state $\ket{1}$ of the oscillator mode. 
Consequently, if the photon is subsequently lost from the oscillator, the qubit will be preferentially projected to the state $\ket{-}$. 
\\
\\
Thus, the steady-state value of $\langle\sx\rangle$ is 0 for case (\textbf{a}) and $-1$ for case (\textbf{b}). Clearly, we see that the qubit dynamics depends on the parameters $\{ E(x),\Delta_y, g,\kb \}$, and in certain cases can have a steady state value of $\langle\sx\rangle$ between 0 and $-1$.

We note here that by considering our full model (Eq. (1) of the main text) with not only dissipation on the coupling mode $\kappa_b$ but also dissipation on the tuning mode $\kappa_a$, we can expect a more complex landscape of steady state dynamics which may qualitatively fall into the above regime.

\subsection{$E(x)=0$}
In this case, we follow the approach in Ref. \cite{Gambetta2008} and begin by writing the density matrix of the qubit and oscillator as
\begin{align}
    \hrho= \ket{g}\bra{g}\otimes \hrho_{gg}+\ket{e}\bra{e}\otimes \hrho_{ee}+\ket{g}\bra{e}\otimes \hrho_{ge}+\ket{e}\bra{g}\otimes \hrho_{eg}
    \label{rhob}
\end{align}
where $\ket{g,e}$ represent the qubit states and $\hrho_{ij}$ acts on the oscillator Hilbert space conditioned on the qubit state. 
From Eq.~\eqref{met},~\eqref{ht} with $E(x)=0$ and Eq.~\eqref{rhob} we get,
\begin{align}
    \dot{\hrho}_{ge}&=-ig(\htb^\dag+\htb){\hrho}_{ge}-ig{\hrho}_{ge}(\htb^\dag+\htb)-i\Delta_y(\htb^\dag\htb {\hrho}_{ge}-{\hrho}_{ge}\htb^\dag\htb)\nonumber\\
    &+\kb \htb{\hrho}_{ge}\htb^\dag-\frac{\kb}{2}\htb^\dag\htb{\hrho}_{ge}-\frac{\kb}{2}{\hrho}_{ge}\htb^\dag\htb
\end{align}
Next, we use the positive-P representation ${\hrho}_{ge}=\int P(\alpha,\alpha^*,t)d\alpha d\alpha^*$ and write the effective Fokker-Planck equation,
\begin{align}
    \frac{\partial P}{\partial t}&=-2igP(\alpha+\alpha^*)+i(g+\Delta_y\alpha)\frac{\partial P}{\partial\alpha}\nonumber\\
    &+i(g-\Delta_y\alpha^*)\frac{\partial P}{\partial\alpha^*}+\kb P+\frac{\kb\alpha}{2}\frac{\partial P}{\partial \alpha}\nonumber\\
    &+\frac{\kb\alpha^*}{2}\frac{\partial P}{\partial \alpha^*}
\end{align}
Now we must solve the above equation with some given initial condition. 
In the setup of interest we start with the oscillator mode in vacuum so that $P(\alpha,\alpha^*,0)=\delta^2(\alpha)=\lim_{\varepsilon\rightarrow 0}(1/\pi \varepsilon)\exp(-|\alpha|^2/\varepsilon)$ and the qubit in the state $\ket{+}$. 
We can make the Gaussian ansatz, $P(\alpha,\alpha^*,t)=\exp(-a(t)+b(t)\alpha+c(t)\alpha^*-d(t)|\alpha|^2)$ and write the equivalent differential equations for $a,b,c,d$ to get,
\begin{align}
    -\dot{a}&=igb+igc+\kb\nonumber\\
    -\dot{d}&=-\kb d\nonumber\\
    \dot{b}&=-2ig+\left(i\Delta_y+\frac{\kb}{2}\right)b-ig d\nonumber\\
    \dot{c}&=-2ig+\left(-i\Delta_y+\frac{\kb}{2}\right)b-ig d
\end{align}

These time-dependent equations can be easily solved with initial conditions now written as $a(0)=\ln \pi\varepsilon,b(0)=0,c(0)=0,d(0)=1/\varepsilon$ but the analytic expressions are considerably simplified when $\Delta_y=0$. 
Once we get $a,b,c,d$ we are able to reconstruct $P$ and hence $\hrho_{ge}$. 
The relevant quantity of interest is the time-dependence of the expectation value of the operator $\ket{g}\bra{e}$,
\begin{align}
    \langle \ket{g}\bra{e} (t) \rangle=Tr[\hrho_{ge}]=\int P(\alpha,\alpha^*,t)d^2\alpha 
\end{align}
For $\Delta_y=0$ and an initial condition where the qubit is in $\ket{+}$ and the oscillator in vacuum, we have
\begin{align}
    \langle \ket{g}\bra{e} (t) \rangle=\frac{1}{2}\exp\left[-\frac{8g^2t}{\kb}+\frac{16g^2}{\kb^2}(1-e^{-\kb t/2})\right]
\end{align}
Next, we consider the rate equation for ${\hrho}_{gg}$,
\begin{align}
     \dot{\hrho}_{gg}&=-ig(\htb^\dag+\htb){\hrho}_{gg}+ig{\hrho}_{gg}(\htb^\dag+\htb)-i\Delta_y(\htb^\dag\htb {\hrho}_{gg}-{\hrho}_{gg}\htb^\dag\htb)\nonumber\\
    &+\kb \htb{\hrho}_{gg}\htb^\dag-\frac{\kb}{2}\htb^\dag\htb{\hrho}_{gg}-\frac{\kb}{2}{\hrho}_{gg}\htb^\dag\htb
\end{align}
Like before, we use the positive-P representation ${\hrho}_{gg}=\int P(\alpha,\alpha^*,t)d\alpha d\alpha^*$, so that
\begin{align}
    \frac{\partial P}{\partial t}&=-ig\frac{\partial P}{\partial\alpha^*}+ig\frac{\partial P}{\partial\alpha}-i\Delta_y\frac{\partial \alpha^*P}{\partial\alpha^*}+i\Delta_y\frac{\partial \alpha P}{\partial\alpha}\nonumber\\
    &+\frac{\kb\alpha}{2}\frac{\partial P}{\partial \alpha}+\frac{\kb\alpha^*}{2}\frac{\partial P}{\partial \alpha^*}
\end{align}
We can again make the Gaussian ansatz $P(\alpha,\alpha^*,t)=\exp(-a(t)+b(t)\alpha+c(t)\alpha^*-d(t)|\alpha|^2)$, and solve the corresponding differential equations for $a,b,c,d$ to get $P(\alpha,\alpha^*,t)=\delta^2(\alpha-\alpha(t))$ where $\alpha(t)=2ig(1-\exp(-\kb t/2))/\kb$ (for $\Delta_y=0$).
Thus, $\langle|g\rangle\langle g|(t)\rangle =1/2$. 
Similarly we can show that $\langle|e\rangle\langle e|(t)\rangle =1/2$. 
Hence the diagonal terms of the qubit density matrix do not decay with time --- only the off-diagonal terms do. 
Thus, in this case the qubit undergoes pure dephasing due to its coupling with the oscillator mode and at steady state becomes an equal mixture of $\ket{g}$ and $\ket{e}$.